\documentclass[aps,prb,twocolumn,superscriptaddress,floatfix,morefloats]{revtex4}
\usepackage{longtable}
\usepackage{subfigure}
\usepackage{amsmath,amssymb,amsfonts,bm}
\usepackage{lscape,graphicx}
\usepackage{pstricks}
\usepackage{ulem}
\usepackage{color}

\def\KeyWord#1{$\backslash$\IfColor{$\!\!$\textRed{#1}\textBlack}{#1}$\!\!$}

\newcommand{\be}{\begin{equation} }
\newcommand{\ee}{\end{equation} }
\newcommand{\ba}{\begin{eqnarray} }
\newcommand{\ea}{\end{eqnarray} }
\newcommand{\n}{\nonumber \\ }

\newcommand{\mac}{\mathcal}

\newcommand{\su}{$SU(2)_k$}
\newcommand{\bit}{\begin{itemize}}
\newcommand{\eit}{\end{itemize}}

\def\em{\it}

\graphicspath{{Figures/}}

\begin{document}

\include{epsfx}

\title{Condensation of achiral simple currents in topological lattice models: a Hamiltonian study of topological symmetry breaking}

\author{F. J. Burnell}
\affiliation{All Souls College, Oxford, United Kingdom}
\author{\!\!\!$^{,2}$\,\,\, Steven H. Simon}
\affiliation{Theoretical Physics, Oxford University, 1 Keble Road, Oxford, OX1 3NP, United Kingdom}
\author{J. K. Slingerland}
\affiliation{Department of Mathematical Physics, National University of Ireland, Maynooth, Ireland}
\affiliation{Dublin Institute for Advanced Studies, School of Theoretical Physics,
10 Burlington Rd, Dublin, Ireland}

\begin{abstract}
We describe a family of phase transitions connecting phases of differing non-trivial topological order by explicitly constructing Hamiltonians of the Levin-Wen[PRB 71, 045110] type which can be tuned between two solvable points, each of which realizes a different topologically ordered phase.  We show that the low-energy degrees of freedom near the phase transition can be mapped onto those of a Potts model, and we discuss the stability of the resulting phase diagram to small perturbations about the model.  We further explain how the excitations in the condensed phase are formed from those in the original topological theory, some of which are split into multiple components by condensation, and we discuss the implications of our results for understanding the nature of general achiral topological phases in $2+1$ dimensions in terms of doubled Chern-Simons theories.
\end{abstract}

\maketitle

\section{Introduction}

The study of non-abelian phases of matter has drawn increasing attention in recent years, inspired in part by potential applications to topological quantum computation\cite{Freedman,Freedman2,KitaevToric,NayakReview}.  Since Moore and Read's\cite{MooreRead} proposal of the Pfaffian wave-function for the fractional quantum Hall state at $\nu = 5/2$, there have been many propositions for realizing non-abelian matter in a wide variety of physical systems.   These include in $\rm Sr_{2}RuO_4$ superconductors\cite{Ivanov,ChungSrRu,DSSrRu,Jang,XiaSrRu,Kidwingira}, Josephson junction arrays\cite{IoffeDucot,IoffeDucot2}, Helium 3A\cite{Volovik,Kopnin}, cold-atom systems\cite{GurariePWave,CooperPWave,CooperReview}, as well as conventional superconductor interfaces on 3D topological insulators\cite{FuKane} and other strongly spin-orbit coupled materials\cite{Masatoshi,SauSO}.  In addition to this wealth of new directions, recent experimental investigations of the $5/2$ state\cite{Willet,Willett0,YacobyCharge,Radu,ShotNoise,shotnoise2,neutralmodes} give renewed incentive to study non-abelian matter.    Despite this boom of interest, one topic which received relatively little attention until recently\cite{TSBPRL,TSBPRL2,TSB,Gilsetal,WenBarkeshli,WenBarkeshliLong,WB2} is the question of phase transitions in these topological systems, which will be the focus of this work.

The non-abelian phases of interest can be characterized by their topological order --- that is, by the {\it fusion} and {\it braiding} (statistical) properties of their low-lying excitations.
An interesting general question is to understand, on broader grounds, the possible transitions between phases with different topological order.
Drawing on the analogy with the Landau theory of symmetry-breaking phase transitions, we consider what features of such transitions, and the relationship between the phases they connect, can be deduced from the topological order alone.   There are two questions to address here.  First, if the phase transition is second-order, one would like to be able to deduce from the topological orders of the initial and final phases  a long-wavelength description of the critical theory.
While we will not provide a complete answer to this question here, we will identify the critical theories for a large class of phase transitions, and comment on their universality.  Second,
we would like to understand -- using information about the topological order alone -- how the excitations of the phases on both sides of the transition are related.  A substantial step in this direction was made by Bais and Slingerland\cite{TSB,TSBPRL,TSBPRL2}, who developed a framework known as {\it topological symmetry breaking}  which describes topological phase transitions as a type of Bose condensation and deduces the topological order of the condensed phase.  Here we will elaborate on this picture, by studying the fate of the excitations of a Hamiltonian which can be tuned through the phase transition.

Our method for addressing the above issues will be to study a class of topological symmetry-breaking phase transitions which we can realize explicitly by  a simple deformation of lattice Hamiltonians of the Levin-Wen\cite{LW} type.  The key advantage of this approach is that, because these Hamiltonians can be solved exactly at certain points in the phase diagram, we will be able to understand in detail the phase transition (which in this case is always of the transverse-field Potts type) and the fate of the excitations on both sides of the phase boundary.  Further, many of the features of these transitions are associated with the topological orders of the phases, rather than to the specific lattice model.  Indeed, from the point of view of studying the general characteristics of the phase diagram, the Levin-Wen models should be viewed not as candidates for accurately describing a theory on the microscopic scale, but as ``$\sigma$-models" for these types of topological phases: though they will differ microscopically from any known candidate physical system, they represent faithfully -- in a highly tractable form-- the long-wavelength dynamics of any gapped phase with the same topological order.  Hence by studying phase transitions in these models, we may hope to extract the generic features of the low-energy theory: the topological orders of the two phases, and (in some cases) the critical theory (or possible critical theories) of the phase transition separating them.  This is exactly as in the Landau symmetry-breaking program.

This work is divided into three main parts.  In Sect. \ref{TransSect}, we will show explicitly how to construct a lattice Hamiltonian which can be tuned between two phases with different topological order.  The transition occurs via condensation of a bosonic plaquette excitation which can be mapped onto a Potts spin; we will show that, under certain assumptions, the critical theory is that of the Transverse-field Potts model (TFPM).  Sect. \ref{TransSect2} discusses the fate of these results when the model is perturbed.   In Sect. \ref{FPhaseSect}, we will study the condensed phase in detail, showing how its excitations arise from those of the uncondensed model.  Of primary interest here is the phenomenon of ``splitting\cite{TSB,TSBPRL2}" (in the context of conformal field theory this is related to fixed point resolution\cite{FPR}) a single species of excitation into two distinct particle types after condensation, which we can understand explicitly in the lattice model.  We conclude with a summary of our results and a brief discussion of other types of transitions which can be studied by the same methodology.

\section{Condensing achiral simple currents} \label{TransSect}

In the present  work we will realize phase transitions in the TFPM universality class by condensing a particular type of boson (an achiral {\it simple current}) which has the property that multiple bosons combine (or {\it fuse}) according to addition rules appropriate to a $Q$-state Potts spin (a $\mathbb{Z}_Q$ spin).
To describe precisely the nature of these transitions, we will study a family of lattice  Hamiltonians that are equivalent to exactly solvable Levin-Wen\cite{LW} models at two points in the phase diagram.  In the uncondensed phase, the topological order will be that of a ``doubled" Chern-Simons theory\footnote{The experts will realize that our results are not restricted to doubled Chern-Simons theory, but apply equally well to any doubled modular tensor category.}-- a chiral Chern-Simons theory together with its mirror image.   Here we will show that when the condensed boson $\Phi$ is a certain type of achiral simple current with $\mathbb{Z}_Q$ symmetry  (i.e., $\Phi^Q \equiv Id$),
the long-wavelength description of the critical theory can be mapped exactly onto the $Q$-state Potts model.  (We will discuss in section \ref{TransSect2} various perturbations to this Potts model).  Condensing achiral currents necessarily leads to a net achiral condensed phase, whose topological order is also captured by a Levin-Wen\cite{LW} model; the study of this phase will be the focus of Sect. \ref{FPhaseSect}.

\subsection{Topological Lattice Hamiltonians}

Our starting point will be an exactly solvable Hamiltonian of the type introduced by Levin and Wen\cite{LW}.  We will restrict our discussion to models which realize doubled Chern-Simons theories, though the construction of Ref. \onlinecite{LW} is more general.

The Levin-Wen models  can be viewed as deformed versions of a lattice Yang-Mills theory.  Their Hilbert space consists of a  finite set of possible states $i \in \{Id, \ldots, r \}$ on each edge of a honeycomb lattice.
The labels $\{Id, \ldots, r \}$ of these states are analogous to the set of possible electric fluxes in the gauge theory.  (The label $Id$ will always denote the trivial flux).
The Levin-Wen Hamiltonian is constructed from two sets of commuting projectors:
\be \label{Eq_HLW}
H =  -\epsilon_e \sum_V  {\cal P}_V - \epsilon_m \sum_P    {\cal P}^{(0)}_P \ \ \ .
\ee
The first sum contains projectors ${\cal P}_V$ acting on vertices $V$, and ensures that the ground states obey the constraints $\nabla \cdot \vec{E} =0$ at each vertex.   For example, if the gauge group is $SU(2)$, in the lattice Yang-Mills theory 
\be
P_V\ \  \includegraphics[totalheight=.4in]{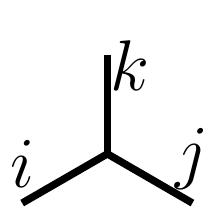} = \sum_{l \in i \times j} \delta_{kl} \ \  \includegraphics[totalheight=.4in]{Vertex.pdf}
\ee
where the rules for addition of angular momenta stipulate that $i \times j = \sum_{l= |i-j|}^{i+j} l$.
 The projectors  ${\cal P}_P^{(0)}$ in the second sum impose  a condition analogous to $\nabla \times B =0$ at each plaquette $P$ on the ground states. (As in lattice Yang-Mills theories, the operator $P_P$ is essentially a superposition of Wilson loop operators encircling the plaquette $P$; we will will return to its the precise form presently.)  All  of the projectors in Eq.~(\ref{Eq_HLW}) commute, since these two constraints can be satisfied simultaneously.

The models are `deformed' in the sense that the number of fields $r$ is finite, even though the gauge group is not discrete.  For example, if the gauge group is $SU(2)$ the lattice Yang-Mills theory would have electric fluxes corresponding to all allowed values $0, 1/2, ... $ of the total spin -- of which there are infinitely many.   In the Chern-Simons theory there is a maximum spin $k/2$; the rules for adding angular momenta which specify $P_V$(or more generally, combining fluxes carrying different representations of the gauge group) must also be modified to be consistent with this truncation.  ($k$ is referred to as the {\it level}, and the resulting Chern-Simons theory is denoted $SU(2)_k$).  Tabulations of these {\it fusion} rules for a number of theories can be found in Refs. \onlinecite{BondersonThesis,BondersonAnnals,KitaevVeryLongPaper}.

It is important here that there is no ``$E^2$'' term in the Levin-Wen Hamiltonian: there is an energy cost to creating matter sources, but once a pair of these is created, there is no Coulomb-like energy cost associated with separating them.
This ensures both that the Hamiltonian is solvable, as all terms commute, and that the ground states and low-lying excitations are independent of all length scales, as required if they are to be topological.

Because all of the projectors in Eq.~(\ref{Eq_HLW}) commute, the topological order of the model can be deduced from the statistical properties of its excitations.
Though the electric fluxes used to label the edges in the lattice are drawn from the representations of a chiral Chern-Simons theory, the Levin-Wen models realize a phase whose topological order is that of a {\it doubled} (net achiral) Chern-Simons theory, whose sources we will label $i_R \times j_L$ with $i,j \in \{Id, .. r \}$ (indicating a composite excitation comprised of a right handed $i$ particle and a left handed $j$ particle).  

Excitations in the Levin-Wen model can be described in terms of electric sources, which violate the vertex condition $P_V =1$, and magnetic sources which violate the plaquette condition $P_P^{(0)} =1$.  As described in Ref. \onlinecite{CH}, these are related to the sources in the doubled Chern-Simons theory as follows.  The electric excitations have the chirality of the original model (assumed to be right handed), so that electric sources are of the form $i_R \times Id_L$ or just $i_R$ for short.   The magnetic sources are achiral composites $i_R \times i_L$ of a particle and its mirror image.  The ``left-handed'' excitations $i_L$ carry both electric charge and magnetic flux of the particle type  $i$.  We will discuss these excitations in more detail when we consider the topological order of the condensed phase in Sect. \ref{FPhaseSect}.

Importantly, while generically electric sources have nontrivial braiding properties, magnetic sources are {\em always} bosons in the sense that they have trivial braiding statistics with themselves, hence can be condensed\cite{TSB}.  (In doubled Chern-Simons theories achiral particles are necessarily bosonic -- meaning that they have trivial self-braiding statistics (in at least one fusion channel)\footnote{In abelian theories, this is because their Berry phase under exchange is the sum of the Berry's phase of their right- and left- handed components -- which are always equal in magnitude and opposite in sign.  More generally, there will be a Berry matrix of phases due to self-braiding in all possible fusion channels.}.)

Here we will condense a particularly simple type of magnetic excitation -- one which behaves essentially like the magnetic flux in a discrete Abelian gauge theory.
Specifically, we choose a particle $\phi$ from the chiral theory with the property that $\phi^Q =Id, \phi^p \neq Id$ for any $p < Q$.   Excitations with this property are known as $\mathbb{Z}_Q$ {\it simple currents}.  In this work we consider condensation of the magnetic excitation $\Phi = \phi_R \times \phi_L$ which is a bosonic  $\mathbb{Z}_Q$ simple current carrying flux $\phi$.
%We emphasize again, that this magnetic excitation $\Phi$ can be thought of as a magnetic flux of $\phi$ penetrating a plaquette, but actually is an achiral particle.
(Throughout this text, we will use $\phi$ to denote the chiral excitation, while $\Phi$ refers to the corresponding achiral magnetic particle.)

The particle we will condense is an excitation which violates only the plaquette projectors of Eq.~(\ref{Eq_HLW}).   To see how to condense such a particle, we must understand in more detail the form of the plaquette projectors.  These have a  very similar form to the plaquette term in a lattice gauge theory: they can basically be viewed as Wilson loop operators around a single plaquette, which in lattice gauge theory contribute a ``$B^2$'' term to the long-wavelength Hamiltonian.  Since we wish to work in a basis where edge labels denote electric flux, the result is  an operator which raises or lowers the electric flux on each edge of the plaquette.  This will be familiar to many readers from the standard formulation of Ising gauge theory, in which the magnetic term in the Hamiltonian is commonly expressed as an operator which simultaneously flips all spins bordering a plaquette.

The amount by which the electric flux changes under these operators is determined by the representation carried by the Wilson line.  To construct a plaquette {\it projector} ${\cal P}^{(j)}$ onto magnetic flux $j$, we must use a particular linear combination of these representations, given by:
\be \label{Eq_Projs}
{\cal P}^{(j)}_P \equiv  \sum_{i=Id}^r  S_{j i }  \hat{W}_{i} (P)\ \ \ .
\ee
where $\hat{W}_{i }(P)$ denotes a Wilson line (or ``string operator" in the language of Ref.~\onlinecite{LW}) around the plaquette $P$ in the representation $i$, and the sum runs over the (finitely many) representations $i$.  Here $S$ is  the {\it modular $S$ matrix} of the {\it chiral} Chern-Simons theory, and is related to the braiding properties of the excitations therein.
The Levin-Wen Hamiltonian (Eq.~\ref{Eq_HLW}) has projectors ${\cal P}^{(0)}_P$ on each plaquette which project onto the identity representation $Id$, in which case
\be \label{Eq_S0}
S_{0 i } = \frac{1}{\mac{D}}\Delta_i
\ee
where $\Delta_i$ is a real number known as the {\it quantum dimension} of the representation $i$, and $\mac{D} = \sqrt{ \sum_{i=Id}^r \Delta_i^2 }$.

More generally,
if $\phi$ is a $\mathbb{Z}_Q$ simple current, then $S_{\phi^n i}$ has the particularly simple form
\be \label{Eq_Smatrix}
S_{\phi^n i } = \frac{1}{\mac{D}} \Delta_i e^{ 2 \pi i\frac{n q_i }{Q} }
\ee
where $q_i$ is specific to $\phi$, but independent of $n$.  (To be precise, $q_i$ is an integer, with $q_i =0$ if $i$ and $\phi$ have trivial mutual braiding statistics in the chiral Chern-Simons theory).  Substituting (\ref{Eq_Smatrix}) into (\ref{Eq_Projs}) gives a projector onto plaquettes with flux $\phi$.

%Since ${\cal P}^{(0)}_P$ is a particular combination of raising operators for the electric flux, its eigenstates are necessarily weighted superpositions over all combinations of electric flux assignments (edge labels) which satisfy the condition $\nabla \cdot \vec{E} =0$ at each vertex.  Information about the energy of such eigenstates is therefore encoded in the relative phases of the elements of this superposition-- for example, in the Ising gauge theory, a state is vortex free if all spin configurations appear with the same sign.  Thus operators which create violations of this projector are precisely the vortex excitations, which introduce extra representation-dependent phases into the ground-state wave function.

\subsection{Condensation}

We now describe the simple deformation of the exactly solvable Hamiltonians which allows us to tune the system through a phase transition in which the boson $\Phi = \phi_R \times \phi_L$ condenses, and derive a second exactly solvable Hamiltonian which captures the topological order of the condensed phase.  Here we restrict ourselves to the case that $\Phi$ is a purely magnetic excitation, and that it is a simple current.

\begin{figure}[h]
\includegraphics[width=2.8in]{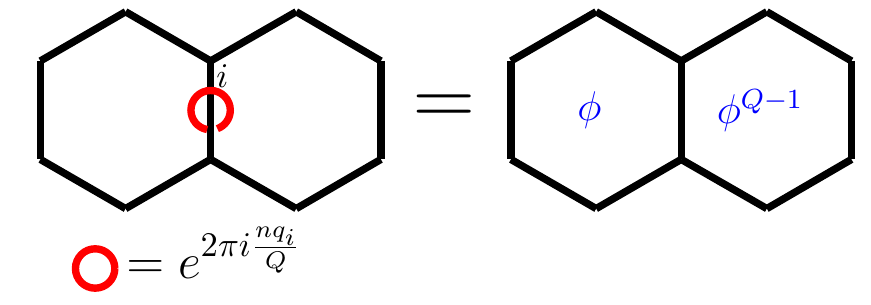}
\caption{ \label{PhiFig}
Creating a pair of simple-current vortices on adjacent plaquettes.  The operator $V^\dag_{e_{12}} (\phi^n) $ acts on the edge separating plaquettes $P_1$ and $P_2$, assigning a phase $e^{ 2 \pi i \frac{n q_i }{Q} } $ to any component of the wave-function in which this edge is labeled by $i$.   If the plaquettes start without flux as shown in the figure, the result is a pair of vortices of flux $\phi^n$ on $P_1$ and $\phi^{Q-n}$ on $P_2$.
 }
\end{figure}

We first require an operator that will pair-create the requisite vortices.
If $\phi$ is a $\mathbb{Z}_Q$ simple current, we may create a pair of vortices of flux $\phi^n$ and $\phi^{Q-n}$ on adjacent plaquettes $P_1$ and $P_2$ by acting with an operator:
\be
V^\dag_{e_{12}} (\phi^n) | i_{12} \rangle   = \frac{1}{\Delta_i } S_{\phi^n i }  |i_{12} \rangle = e^{ 2 \pi i \frac{n q_i }{Q} }  |i_{12} \rangle
\ee
where $e_{12}$ is the edge between plaquettes $P_1$ and $P_2$, and
$| i_{12} \rangle $ denotes any state that has electric flux $i$ on the edge $e_{12}$ shared by plaquettes $P_1$ and $P_2$.
(Which plaquette obtains a vortex of flux $\phi^n$, and which a flux $\phi^{Q-n}$, depends on which orientation we choose for the electric flux on the edge $i$.)  This operator $V^\dag_{e_{12}} (\phi^n) $ thus assigns a phase to the wave-function depending on the value of the electric flux on the edge separating plaquettes $P_1$ and $P_2$ (Fig. \ref{PhiFig}).  Note that if the plaquettes already have magnetic fluxes $\phi^a$ and $\phi^b$, then this operator simply increments these fluxes mod $Q$ accordingly to $\phi^{(n+a) \rm{mod} Q}$ and $\phi^{(Q-n+b) \rm{mod} Q}$.

%\begin{center}
%\end{center}

To engender condensation, we will simultaneously decrease the energy gap of $\Phi^n$ excitations  (magnetic fluxes of $\phi^n$) without changing the gap to the model's other excitations, and also add a term  proportional to $V^\dag (\phi)$ which creates such excitations spontaneously.
The Hamiltonian that does this has the form:
\ba \label{Eq_HFull}
H &=&  -\epsilon_e \sum_V  {\cal P}_V -\frac{ \epsilon_m }{2} \sum_P    \sum_{n=0}^{Q-1}  {\cal P}^{(\phi^n)}_P    \\
&& - \frac{\epsilon_m}{2} \alpha_T \sum_P    \left[ {\cal P}_P^{(0)} -  \sum_{n=1}^{Q-1} {\cal P}^{( \phi^n)}_P  \right ] - \epsilon_m \alpha_N  \sum_{e_{ij}} \hat{V}^\dag_{e_{ij}}  \nonumber \ \ \ .
\ea
(Note that the sums over powers of $\phi$ in the first and second lines run from $0$ to $Q-1$ and $1$ to $Q-1$, respectively). 
Here $e_{ij}$ is the edge shared by plaquettes $P_i$ and $P_j$, and %hence the final sum is over all edges,
\be \label{Eq_V}
\hat{V}^\dag_{e_{ij}}  =\frac{1}{Q} \sum_{n=0}^{Q-1} V^\dag_{e_{ij} } (\phi^n)
\ee
creates all fluxes $0,\phi,\phi^2, \ldots, \phi^{Q-1}$ with equal amplitude.    Note that here $\hat V^\dagger_{e_{ij}} = \hat V_{e_{ij}} = \hat V_{e_{ji}}$.

In the limit $\alpha_T =1, \alpha_N =0$, this Hamiltonian is just the exactly solvable Levin-Wen Hamiltonian (Eq.~\ref{Eq_HLW}), which assigns a mass gap of $\epsilon_e$ to electric sources, and $\epsilon_m$ to magnetic sources.
The second line of Eq.~\ref{Eq_HFull}  allows us to tune the model through a condensation, by varying the parameters $\alpha_N$ and $\alpha_T$.

To understand the phase portrait of (\ref{Eq_HFull}), it is helpful to consider the effect of the various terms on the subspace of states in which the only excitations are the magnetic $\Phi$ excitations which we wish to condense.
The first line of (\ref{Eq_HFull}) is indifferent to the presence of $\Phi$, as $\Phi$ does not violate the vertex condition, and the sum of plaquette projectors
\be
\sum_{n=0}^{Q-1}  {\cal P}^{(\phi^n)}_P
\ee
has eigenvalue $1$ for a plaquette excitation $\Phi^n$ (magnetic flux $\phi^n$, for $n=0,1,... Q-1$), and $0$ for a plaquette containing any other possible excitation.   The role of these terms of the Hamiltonian is to ensure that there is a minimum gap of $\frac{\epsilon_m}{2}$ for all magnetic excitations other than $\Phi^n$, and $\epsilon_V$ for all electric excitations, everywhere in the phase diagram.  (Here we keep $\epsilon_m$ and $\epsilon_V$ fixed).  These energy gaps are unfrustrated, in the sense that neither the plaquette term nor the creation term on the second line have any amplitude to create these other excitations, which therefore remain both gapped and orthogonal to the ground state throughout the phase transition we describe.   Thus we will be able to focus our entire attention on the Hilbert space containing only by $\Phi^n$ type excitations.

The second line of (\ref{Eq_HFull}) contains the terms which drive the system through a condensation transition.  The first term
\be \label{Eq_PottsPl}
 - \frac{\epsilon_m}{2} \alpha_T \sum_P    \left[ {\cal P}_P^{(0)} -  \sum_{n=1}^{Q-1} {\cal P}^{( \phi^n)}_P  \right ]
 \ee
sets the scale of the gap to creating a $\Phi$ excitations, which we may tune from $\epsilon_m$ at $\alpha_T =1$ to $0$ at $\alpha_T =0$.  For $\alpha_T<0$ the formation of $\Phi$ excitations is energetically favored; for small negative $\alpha_T$ the system remains in the condensed topological phase.  If, however, we make $\alpha_T<0, |\alpha_T| \gg \alpha_N$ we will find a new variant of the uncondensed phase in which each plaquette is occupied by a definite superposition of $\phi^n$ vortices.  (Taking $\alpha_N=0$ in this phase gives another exactly solvable model, again of the general form discussed in Ref. \onlinecite{LW}).  

The term in  (\ref{Eq_HFull}) with coefficient $\alpha_N$ adds an amplitude to create or destroy pairs of fluxes on adjacent plaquettes (as shown in Fig.~\ref{PhiFig}).  Analogous to adding an anomalous term to a Hamiltonian of conventional bosons
\be
  H = H_0 + \Delta^* \, b^\dagger  + \Delta b,
\ee
spontaneous creation of $\Phi$ particles is precisely what we should expect to need to add to a Hamiltonian in order to form a $\Phi$ condensate.  In these topological models, particles must be created in particle-antiparticle pairs connected by a Wilson line (or {\it string operator}); $\hat{V}^\dag_{e_{ij} } $ acts  on pairs of neighboring plaquettes to create these.

It is instructive to consider the Hamiltonian (\ref{Eq_HFull}) in the limit $\alpha_T = 0, \alpha_N =1$.  The plaquette term then has the form:
\ba
\frac{\epsilon_m}{2} \sum_{n=0}^{Q-1} {\cal P}_P^{(\phi^n)}  &= &\frac{\epsilon_m}{2}  \sum_{j=Id}^r \Delta_j \left( \sum_{n=0}^{Q-1} e^{2 \pi i  \frac{n q_j}{Q} } \right ) \hat{W}_j(P)  \n
&=& Q\frac{\epsilon_m}{2}   \sum_{j=Id}^r  \delta_{q_j, 0} \Delta_j \hat{W}_j (P)
\ea
where $\hat{W}_j (P)$ is the Wilson line (or string operator) around plaquette $P$ carrying the representation $j$.  For representations which have trivial braiding statistics with the $\phi$ flux, $q_j =0$, and the second sum on the right hand side (in the parentheses) just gives an overall factor of $Q$.  If $q_j \neq 0$ (mod $Q$), then the second sum leads to complete destructive interference, and the Wilson line $j$ is eliminated from the plaquette term.
 %The resulting plaquette operator may still in general change the labels $i$ on edges surrounding the plaquette, but will not change $q_i$,  since if $q_i  =0$, then $\hat{W}_i |j\rangle = \sum_{k} \alpha_k |k \rangle$ where $q_k= q_j$.

Similarly, the effect of the vortex creation term $\hat{V}$ on the edge label $i$ on edge $e$ is:
\be
\frac{1}{Q} \sum_{n=0}^{Q-1} V^\dag_{e} (\phi^n) | i \rangle = \frac{1}{Q}\sum_{n=0}^{Q-1} e^{2 \pi i  \frac{n q_i}{Q} }   | i \rangle =  \delta_{q_i, 0} |i \rangle \ \ \ .
\ee
Hence $-\alpha_N \epsilon_m \hat{V_e}$ effectively assigns an energy cost $\alpha_N \epsilon_m$ to any (electric flux) label $i$ on edge $e$ for which $q_i \neq 0$.  This creates a confining potential for any label which braids non-trivially with the flux $\phi$ of the condensed boson $\Phi=\phi_R \times \phi_L$.

In addition to the Levin-Wen point $\alpha_T=1, \alpha_N=0$, we therefore have a second special point in the phase diagram at $\alpha_T =0, \alpha_N >0$ where the Hamiltonian is again exactly solvable.  This point represents the exactly solvable (and fully topological) limit of the condensed phase, in which labels with $q_i \neq 0$ (which do not braid trivially with $\phi$) have been completely eliminated from the theory.   This can be consistently done because the plaquette term (which now contains only raising operators with $q_i =0$) does not mix edge labels with different $q_i$.  Hence states containing only edge labels with $q_i = 0$ comprise the entire low energy space of the Hilbert space in this limit.

Restricted to these states, at $\alpha_T = 0, \alpha_N =1$, the Hamiltonian (\ref{Eq_HFull}) has a particularly simple form.   Since $\hat{V}_e \equiv 1$ on all remaining states, we may drop it from the Hamiltonian, leaving
\be \label{Eq_HEffC}
H_{eff} = -\epsilon_e \sum_V {\cal P}_V - Q \frac{\epsilon_m}{2} \sum _P \tilde{{\cal P}}_P^{(0)}
\ee
where
\be
\tilde{{\cal P}}_P^{(0)} \equiv \sum_{q_i = 0 }  \Delta_i \hat{W}_i (P)
\ee
has the general form of a plaquette projector onto $0$ flux, as per Eq.~(\ref{Eq_Projs}).
As $\tilde{{\cal P}}_P^{(0)}$ and ${\cal P}_V$ commute, Eq.~(\ref{Eq_HEffC}) is again a Hamiltonian comprised of commuting projectors, of the same general form as our initial Levin-Wen Hamiltonian.  The difference here is that the labels $i$ for the edges are now
drawn from the subset of the original labels for which $q_i =0$.  We will study the consequences of this restriction on the model's excitations in Sect  \ref{FPhaseSect}.

\subsection{Effective Potts model of the phase transition}

We have thus argued that (\ref{Eq_HFull})  can be tuned through a phase transition between two solvable Hamiltonians of the Levin-Wen form, with each of the solvable models capturing exactly the topological features of one of the two phases.  (See Ref. \onlinecite{Bombin} for a discussion of an analogous family of achiral solvable lattice Hamiltonians inspired by Kitaev's toric code models\cite{KitaevToric}).  Here we will exploit a natural mapping between the long-wavelength dynamics of (\ref{Eq_HFull}) and the Potts model to identify the phase transition as that of the transverse-field Potts model.

To carry out this program, we define an effective low-energy Hilbert space which we will map exactly onto the Hilbert space of the Potts model.  Specifically, since all excitations other than $\Phi^n$ remain gapped (with a minimum energy of min$\{ \epsilon_V, \frac{\epsilon_m}{2} \}$), we need only consider the subspace of states containing just the ground states of the initial Levin-Wen model, and states which can be derived from these by adding some number of magnetic $\Phi$ (and $\Phi^n$) excitations.  We emphasize that this subset is closed under the dynamics of Eq.~(\ref{Eq_HFull})  for all parameter values of the Hamiltonian, and this subspace contains all modes which become critical at the phase transition.  Hence this low-energy subspace can be consistently separated from the rest of the Hilbert space in order to understand the dynamics of the critical point.

\begin{center}
 \begin{figure}[ht]
 \includegraphics[width=3.3in]{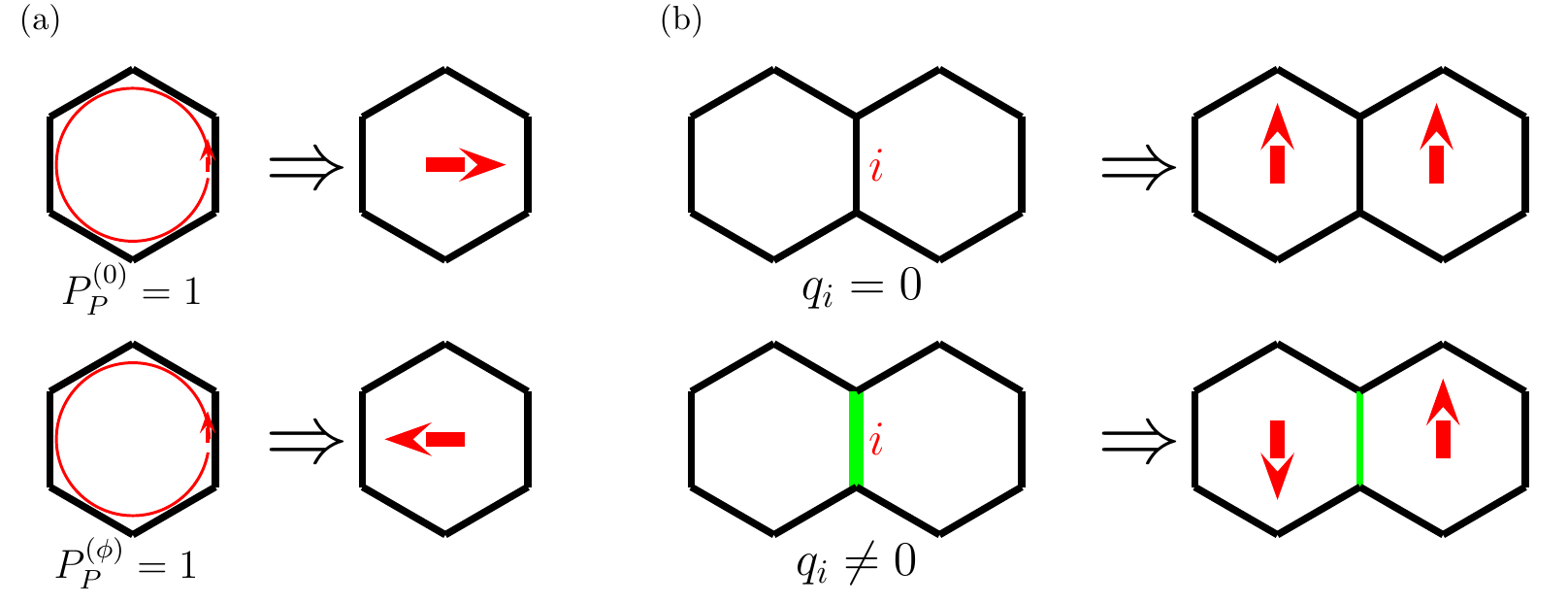}
 \caption{ \label{PottsFig} Effective mapping from the low-energy sector of the string-net model to the Potts model, illustrated here for the Ising case.  (a) In the low-energy sector, we retain only magnetic excitations of flux $\phi^n$, as all other excitations remain gapped throughout the phase transition.  Up to an index specifying the topological ground state sector, the relevant states are specified uniquely by assigning a Potts spin $n \in \{ 0 . . . Q -1 \}$ to each plaquette.   In this figure, an arrow pointing right (left) indicates $n=0$ ($1$) on that plaquette.  (In the spin model we identify these with $S_x = 1$ and $-1$ respectively.)  (b) The terms in the second line of the Hamiltonian Eq.~(\ref{Eq_HFull}) act non-trivially on these states: the plaquette term controls the energetic cost of creating a flux, which we identify with the Potts transverse field.  The flux-creation term $\hat{V}_e$ gives a ferromagnetic Potts interaction in the basis of Eq.~(\ref{Eq_PottsBasis}), indicated here by arrows pointing up  (down) to denote the states $l=0$ ($l=1$) on that plaquette.  (or $S_z = 1$ and $-1$). The eigenvalue of $\hat{V}_e$ is $1$ if the edge $e$ is labeled by a representation $i$ with $q_i =0$, and $0$ otherwise -- indicating that any edge label $i$ with $q_i \neq 0$ (mod $Q$) signals a domain wall in the Potts model.  }
 \end{figure}
 \end{center}

Because there is a unique way to combine such excitations ($\Phi^a \times \Phi^b = \Phi^{(a +  b)  {\rm mod} Q}$), a state in the Hilbert space is uniquely determined by specifying the topological ground state sector and the number $n_P$ of $\phi$ fluxes on each plaquette $P$.
%\footnote{This uniqueness is not generically the case -- for example, if the representation $\Phi$ satisfied Fibbonacci fusion rules ($\Phi \times \Phi = 1 + \Phi$), then to uniquely identify the state with $N$ plaquette violations we would need to know not only their positions, but also the outcome of $N$ pair-wise fusions.}.
Hence we may define the basis
\be \label{Eq_Nbasis}
|g;  n_1, n_2, ... n_N \rangle
\ee
where and $n_i \in 0,\ldots, Q-1$ denotes the number of $\phi$ flux quanta through plaquette $i$  (i.e, the plaquette contains excitation $\Phi^{n_i}$).    Here, $g$ labels a ground state sector of the unperturbed Levin-Wen model (Eq.~\ref{Eq_HLW}) in cases where the manifold has handles.   Our mapping to this $Q$-state model on plaquettes (i.e., on the dual lattice) is illustrated in Fig.~\ref{PottsFig} a.

Readers familiar with the general theory of TQFT's should note that a state of the form (\ref{Eq_Nbasis}) {\it uniquely} identifies a state in the original lattice model because $\Phi$ is a simple current.  For other types of magnetic excitations, such as Fibonacci anyons\cite{Gilsetal}, to specify the state of the lattice model uniquely requires additional information about the relative fusion channels of the excitations.

The terms in the first line of (\ref{Eq_HFull}) act as the identity on our effective Hilbert space. In the basis (\ref{Eq_Nbasis}), the two terms in the second line have the matrix elements:
\ba \label{Eq_PlMat1}
\langle n'_1 |{\cal P}_{P_1}^{(0)} -  \sum_{n=1}^{Q-1} {\cal P}^{( \phi^n)}_{P_1} |  n_1 \rangle = \delta_{n_1', n_1} \left( 2 \delta_{n_1, 0} - 1 \right )  \\
  \label{Eq_PlMat2}
\langle n'_1, n'_2 | V^\dag_{e_{12}} (\phi^k) | n_1, n_2 \rangle = \delta_{n'_1 \pm k, n_1} \delta_{ n'_2 \mp k, n_2} \ \ \
\ea
with the $\pm$ in the second line given by the chosen orientation of the bond $e_{12}$.   In this reduced Hilbert space, below the gap to excitations out of this Hilbert space our Hamiltonian Eq.~\ref{Eq_HFull} is then {\it precisely}
\ba
 H_{\mbox{spin}} &=& -\frac{\epsilon_m}{2}  \alpha_T \sum_i  \sum_{n_i}   (2 \delta_{n_i,0} - 1)|n_i\rangle\langle n_i| \\ &-&\epsilon_m \alpha_N \sum_{<i,j>} \sum_{n_i, n_j} \sum_k   |n_i + k, n_j-k\rangle\langle n_i, n_j|  \nonumber
\ea
The Hamiltonian (\ref{Eq_HFull}) keeps the gap to all powers $n = 1, \ldots Q-1$ of $\Phi$ the same, reflected in the first term in $H_{\mbox{spin} }$.   The second term gives equal amplitudes for transitions that increment and decrement fluxes by any $k$, in accordance with the creation operator (\ref{Eq_V}).    These choices will give us precisely an effective $Q$-state transverse-field Potts model (TFPM) of the phase transition.   In section \ref{sub:within} below we will consider how we may change modify $H_{\mbox{spin}}$ and obtain many possible Hamiltonians for $\mathbb{Z}_Q$ Potts-like spins (for example, the $Q$-state clock model), and discuss how this may affect the phase diagram.
%For example, one could chose an $n$-dependent gap and orchestrate a transition in the universality class of the clock model.

To obtain a more convenient representation of the TFPM, we re-express (\ref{Eq_PlMat1}), (\ref{Eq_PlMat2}) in the basis
\be \label{Eq_PottsBasis}
| l \rangle = \frac{1}{\sqrt{Q} } \sum_{n=0}^{Q-1} e^{ 2 \pi i \frac{l n}{Q} } | n \rangle  \ \ \ .
\ee
The matrix elements (\ref{Eq_PlMat2}) become precisely the Potts interaction:
\be
\langle l'_1, l'_2 | \hat V^\dag_{e_{12}}| l_1, l_2 \rangle = \delta_{l'_1, l_1} \delta_{l_2', l_2} \delta_{l_1 -l_2, 0}
\ee
while (\ref{Eq_PlMat1}) becomes the appropriate transverse-field term\cite{SenthilMajumdar}:
\be
\langle l'_1 |{\cal P}_{P_1}^{(0)} -  \sum_{n=1}^{Q-1} {\cal P}^{( \phi^n)}_{P_1} |  l_1 \rangle =  \frac{2}{Q} -\delta_{l l'}
 \ \ \ .
\ee
This maps the long-wavelength description of the phase transition precisely onto the conventional representation of the transverse-field Potts Hamiltonian:
\ba \label{Eq_HPotts}
H_{\mbox{spin}}&=&  - \frac{\epsilon_m}{2} \alpha_T \sum_{i, l'_i, l_i} \left( \frac{2}{Q}|l'_i \rangle \langle l_i | - | l_i \rangle \langle l_i | \right )  \n
&& - \epsilon_m \alpha_N \sum_{<ij>} \delta_{l_i, l_j} |l_i,l_j  \rangle \langle l_i,l_j  |\ \ \ .
\ea
(Our choice of notation now becomes clear: $\alpha_T$ is the transverse field term and $\alpha_N$ is the neighbor interaction).  Here $\alpha_N >0$ is used to promote condensation, so the model has a  ferromagnetic phase in which all of the $l$-spins are aligned and a paramagnetic phase where they align with the transverse field.

In terms of the Potts model description, the labels which become confined in the condensed phase correspond to domain walls of the ferromagnetic Potts model.  Specifically, if the edge $e_{ij}$ separating plaquettes $i$ and $j$ is labeled $i_{ij}$, then the Potts interaction is
\be
\hat{V}_{e_{ij}} | i_{ij} \rangle = \delta_{q_{e_{ij}}, 0} | i_{ij} \rangle \ \ \ \ \equiv \delta_{l_i - l_j, 0} | l_i, l_j \rangle \ \ \ .
\ee
Hence if $q_{e_{ij} } \neq 0$, then in the basis of Eq.~(\ref{Eq_PottsBasis}), $l_i - l_j \neq 0$ and there is a domain wall in the ferromagnetic Potts representation.  Thus these labels become confined since domain walls are confined in the ferromagnetic phase  (See Fig.~\ref{PottsFig}.b).  (Though the full Hilbert space of the lattice model also includes open strings, for $\epsilon_V>0$ these do not occur in the zero-temperature phase transition, due to the fact that all terms in $H$ commute with $P_V$).

There is one important difference between the lattice model and its spin analogue: while the latter has $Q$ symmetry-related ferromagnetic ground states, the former has only one.  
The topological model is only sensitive to the locations of the domain walls in the Potts model, not to the orientation of the Potts spins in the basis (\ref{Eq_PottsBasis}).  (This loss of ground-state degeneracy also occurs in the dual loop-gas representation of the Ising model).  
This is because the Hilbert space of (\ref{Eq_HFull}) has the restriction that
\be
  \sum_{i=1}^N n_i = 0 \mbox{  mod  } Q.
\ee
That is, the Hilbert space we use is not identical to that of the Potts model; rather, since vortices can only be created in vortex-anti-vortex pairs in the topological theory, it contains only states connected to the vacuum by the action of the Potts interaction (or pair-wise spin flips, in the Ising case).  Since the spin Hamiltonian connects only states within this subspace, this difference does not affect the system's dynamics or the nature of the phase transition.  
However, the $Q$-fold the degeneracy of ferromagnetic ground states in the Potts model is absent in the topological model: in the latter, there is no operator which measures the eigenvalue of $l$ on a particular plaquette, since such an operator would necessarily change the number of vortices on only one plaquette.  We note that this difference renders certain interesting modifications of these models\cite{KitaevPhase,VidalArxiv} difficult to realize in the present context.

We may therefore conclude that the phase transition of (\ref{Eq_HFull}) is precisely that of a ferromagnetic transverse-field Potts model in $2+1$ dimensions, since we have constructed an explicit mapping between the two.   The original string-net model maps to the paramagnetic phase $\alpha_T  \gg \alpha_N$, where the transverse field dominates to ensure that vortices are relatively rare.  In the ferromagnetic phase of the Potts model, the vortex creation term dominates in Eq.~(\ref{Eq_HFull}).   In
this r{\'e}gime $\Phi^n$ vortices have condensed and certain edge labels become confined.  The transition between the two will be in the 2+1D transverse-field-Potts universality class (which is first order for $Q\geq 3$\cite{PottsFirstO,PottsFirstO2}).

The notion that a phase transition which changes the topological order can be described by a spin model, in which a global symmetry is broken when a local variable obtains an expectation value, is somewhat counter-intuitive.
To understand this correspondence, notice first that after projecting onto the states relevant to the critical theory, we arrive at an effective description in a pure (matter-free) discrete Abelian gauge theory.  That is, the vortices that we condense behave exactly like vortices of a discrete Abelian gauge theory, both in terms of how multiple vortices combine, and from the way in which the operators which create them act on the gauge-invariant states.  The mapping to the spin model then simply exploits the fact that this discrete Abelian gauge theory without matter sources is dual to a spin model\cite{Wegner,FradkinShenker} (a model of matter without magnetic vortices).  This duality has been exploited previously to study phase transitions in the Toric code\cite{TrebstTC,Lidar}.  Once matter sources are added to the gauge theory, the dual theory is again a discrete gauge theory: since itinerant charges will always feel the Berry's phase of the condensing vortices, we must include gauge fields in the dual spin model such that the true gauge-invariant order parameter is non-local in both representations.

\subsection{Example}

To make the discussion of the previous section concrete, let us give here an example.  Other, more general examples are discussed in Sect. \ref{ExampleSect}.

First, as described in Ref. \onlinecite{TSBShortP}, we may construct the initial Hamiltonian from representations of the Chern-Simons theory $SU(2)_2$.  In this case there are three possible values $ 0( \equiv Id), 1/2,$ and $1$ of the total spin.  The (commutative and associative) fusion rules of these spins are similar to angular momentum addition, except that the rules are truncated such that no value greater than 1 is ever obtained:
\ba    \label{Eq_Fuse} \nonumber
  0 \times j &=& j \\ \nonumber
  1/2 \times 1/2 &=& 0 + 1  \\  \nonumber 1/2 \times 1 &=& 1/2  \\ 1 \times 1 &=& 0
\ea
Note that the particle $1$ here is a $\mathbb{Z}_2$ simple current.

The vertex projectors of the Levin-Wen model, based on these fusion rules, have eigenvalue $1$ if the edges incident on a vertex are in one of the following $3$ combinations (here the order in which they appear is not important):
\be
(0,0,0) \ \ \ (0, 1/2, 1/2) \ \ \ (0,1,1) \ \ \ (1/2, 1/2, 1) \ \ \ ,
\ee
and the vertex projector gives zero for any other combination of edge labels.

The Wilson lines (whose action is also based on the rules (\ref{Eq_Fuse} )) raise and lower the edge labels according to:
\ba \label{Eq_SU22}
\hat{W}_0 | i \rangle = |i \rangle \ \ \ \ \   \hat{W}_{i } | 0 \rangle = |i \rangle \n
\hat{W}_{1/2} | 1/2 \rangle \propto \alpha |0 \rangle + \beta | 1 \rangle \ \ \ \ \  \hat{W}_{1/2} | 1 \rangle \propto |1/2 \rangle \n
\hat{W}_{1} | 1/2 \rangle \propto |1/2 \rangle \ \ \ \ \hat{W}_{1} | 1 \rangle \propto |0 \rangle
\ea
Here the constants of proportionality depend in general on the labels of adjacent edges, as well as the edge being acted upon (See Ref. \onlinecite{LW} for details); however, their precise value is unimportant for our purposes.  The last line reflects the fact that the spin $1$ particle is an order $2$   ($\mathbb{Z}_2$) simple current: raising the edge label $1$ by $\hat{W}_1$  necessarily gives the trivial label $0$.

We condense the spin $1$ magnetic excitation ($1_R \times 1_L$), which is a $\mathbb{Z}_2$ bosonic simple current.  As described in the previous section, the effective theory below the gap can be mapped precisely onto that of a $2$d transverse-field Ising model.  To do so, we assign a spin variable $S^x = -1$ to every plaquette containing a spin-$1$ vortex, and $S^x =1$ to vortex-free plaquettes.  The relevant components of the $S$ matrix have the form:
\be
S_{1, 0 } = S_{ 1,  1} = 1 \ \ \ S_{0,  \frac{1}{2} } =  \sqrt{2}  \ \ \ S_{1,  \frac{1}{2} } = - \sqrt{2}  \ \ \ .
\ee
In particular, we have:
\ba
\frac{1}{2} \left( {\cal P}_P^{(0)} + {\cal P}_P^{(1)} \right)& =& 1 + \hat{W}_1 \label{Eq_P1} \\
\frac{1}{2} \left( {\cal P}_P^{(0)} - {\cal P}_P^{(1)} \right) &=& \sqrt{2} \hat{W}_{1/2}
\ea
and the term which creates a pair of vortices (and hence flips the spins) on adjacent plaquettes is
\be
\hat{V}_{e_{ij}} = (-1)^{2 \hat{s}_{e_{ij}}} \equiv S^z_i S^z_j
\ee
where $\hat{s}$ measures the total spin of the representation on the edge $e_{ij}$.
Thus we identify the spin-$1/2$ label with the domain wall in the Ising model: an edge carrying the spin-$1/2$ representation necessarily separates two plaquettes with oppositely oriented Ising spins.

In the solvable limit of the condensed phase, we are left with only the edge labels $0$ and $1$.  The vertex condition is now that an even number of edges labeled $1$ must enter each vertex, while the plaquette  projector (\ref{Eq_P1}) flips the label on all edges surrounding the plaquette from $0$ to $1$ or vice versa.   Some readers may recognize this as the Hamiltonian for Kitaev's Toric code (or Ising gauge theory with matter).  We will return to this point in Sect. \ref{FPhaseSect}.

\section{Universality in topological symmetry breaking transitions} \label{TransSect2}

Thus far, we have established a precise mapping between a family of Hamiltonians with the special property that they are exactly solvable at two points in the phase diagram, and (at energies below the minimum gap to excitations at these solvable points) the transverse-field Potts model.   However, our choice of Hamiltonian (\ref{Eq_HFull}) is quite non-generic: we would expect that any real physical system whose long-wavelength dynamics are described by the lattice model would include perturbations away from the solvable Levin-Wen limit everywhere in the phase diagram.  Hence to apply our understanding of the phase diagram of the topological lattice model to more general systems, we must understand the effect of generic perturbations on its behavior.

Ideally, we would like to achieve a framework analogous to the Landau approach to symmetry-breaking phase transitions.  There are two important elements to this analogy.  First, a phase can change its symmetry only by a phase transition.  Specifically, in a gapped system small perturbations to the Hamiltonian which do not close the gap necessarily leave the symmetries intact.  (The exceptions to this rule are systems which undergo first order transitions, or second-order transitions associated with the formation of microscopic domains\cite{LaumannSondhi}, in which case the gap itself need not close, but there are other singularities associated with the phase boundary).
Second, near a second-order phase transition the long-wavelength behavior is largely determined only by the symmetry being broken.  Specifically, symmetry dictates the nature of the field theory at the transition.   There can be specific choices of parameters for which not all relevant operators compatible with the symmetries of the two phases appear in the critical theory; however, these non-generic systems lie at unstable critical points which flow to different critical theories if the perturbation is added.

We emphasize that though the present work treats transitions in which a simple current (or Potts spin) condenses, these questions are relevant to more general TSB transitions.   It is also possible to deform Levin-Wen Hamiltonians (by adding appropriate analogues of $\hat{V}_e$, and modifying the plaquette projectors appropriately) to condense other types of bosonic vortices, such as the Fibbonaci anyons for which a transition of this type has been discussed by Ref. \onlinecite{Gilsetal}.  This can be done without closing the gap to electric sources, so that again we may imagine studying an effective ``spin model" which describes only the ground states and condensing vortices.  (The resulting spin models will not be Potts-like, however).  However, not all such constructions yield a second set of parameter values (analogous to $\alpha_T =0, \alpha_N =1$) at which the Hamiltonian is again exactly solvable; hence it is more difficult to establish the properties (or existence) of the condensed phase.  To grasp completely the phase portrait of achiral topological phases would require an understanding of these more complex transitions and their universality, which we do not undertake here.

 The  notion that a system cannot change its symmetry without undergoing a phase transition has a well-understood analogue for topologically ordered phases.
Specifically, it is known\cite{Hastings,Hastings2} that any deformation to a topologically ordered phase which does not close the quasi-particle gap cannot change the topological order.   Working within the Hilbert space of the model at hand, this means that we may include small perturbations which admix some number of electric or magnetic sources with the ground states, which nonetheless do not change the topological ground state degeneracy or long-range braiding statistics that we associate with the topological order of the solvable Hamiltonian.  Hence topological order in the present context plays the role of symmetry in the more familiar Landau paradigm.  (Since topological orders can often be associated with gauge theories, this is very natural.  The symmetry broken at the phase transition, in these cases, is simply a gauge symmetry rather than a global one).

The second question -- of whether the critical theory is robust against such perturbations -- is less well understood in the topological context.  In the previous section we chose a special trajectory through the phase transition in which the degrees of freedom implicated in the phase transition could be mapped exactly to
those of a transverse-field Potts model.  This suggested that the  critical theory is dictated by the nature of the condensing boson.  Specifically, when this obeys $\mathbb{Z}_Q$ fusion rules, we expect to find critical theories in the universality class of a Q-state spin model  (a Potts model or similar).
To understand whether this conclusion about the critical theory is valid more generally, we must consider the effect of generic perturbations to the Hamiltonian (\ref{Eq_HFull}).

It is convenient to separate the possible perturbations into two classes.  For $Q>3$, it is possible to introduce a special type of perturbation which  changes the parameters of the Potts Hamiltonian (\ref{Eq_HPotts}).  As we explain below, if $\mathbb{Z}_Q$ contains proper subgroups (i.e., for $Q$ not prime), this results in a richer phase diagram than the one described above.  Other types of perturbations may not have analogues in the Potts description; our chief concern here we will be to consider their effect on the critical theory perturbatively.

\subsection{Perturbations within the Potts model subspace}
\label{sub:within}

Let us begin by considering perturbations whose effect is to change the relative strengths of the ferromagnetic couplings or transverse-field terms for the various powers of $\phi$ in Eq.~(\ref{Eq_HFull}).  For $Q>3$ this allows considerable scope to modify the spin model, leading to a rich phase diagram.   A thorough understanding of the behavior of these more general spin models is in itself an interesting question, whose general features we will outline here by reference to the existing literature where possible.

We consider perturbations of the form
\ba     \label{Eq_HPert1}
\delta H &=& \sum_P \delta H_P - \sum_{e_{ij} } \delta H_{e_{ij} } \\
 \delta H_P &=&  \sum_{n=0}^{Q-1} \epsilon^{(P)}_n {\cal P}_P^{(\phi^n)}  \ \ \  \delta H_{e_{ij} }= \frac{1}{Q} \sum_{n=0}^{Q-1} \epsilon^{(e)}_n V^\dag_{e_{ij} } \left (\phi^n \right ) \nonumber
\ea
where $\epsilon^{(P, e)}_n$ can be chosen arbitrarily provided that: 1) $\epsilon^{(e)}_n = \epsilon^{( e)}_{Q-n}$ (required for hermiticity); 2)  the sign of the net transverse field term for any power of $\phi$ remains positive; and 3) that all Potts interactions remain ferromagnetic.  Provided these three conditions are met, the perturbed Hamiltonian will still undergo (one or more) TSB-type phase transitions.  Here we will  consider the case where $\epsilon_n^{(P)}$ and $\epsilon_n^{(e)}$ are real, though the complex case has also been studied in the context of spin models\cite{ChiralPotts}.

Since the perturbations  (\ref{Eq_HPert1}) clearly have no impact on the mapping to $Q$-state spins, we may consider the effect of these perturbations within the spin picture. In terms of the spin state $|n_P\rangle$ on the dual lattice (recall that a state $|n_P\rangle$ indicates a vortex particle $\Phi^n$ on the plaquette $P$ in the original model), the operators in Eq.~(\ref{Eq_HPert1}) are mapped to the operators:
\ba \nonumber
  \delta H_{\mbox{spin} } &=& \frac{1}{Q} \sum_{<i, j>} \, \, \sum_{n_i,n_j,n=0}^{Q-1} \epsilon^{(e)}_n |n_i-n, n_j+n  \rangle \langle n_i, n_j |
\\  &&+ \sum_i \sum_{n_i=0}^{Q-1} \epsilon^{(P)}_{n_i} |n_i \rangle \langle n_i |  \,\,\, .
\ea
The first term flips pairs of spins on neighboring plaquettes; the second is the (transverse) magnetic field.
To make the correspondence to the Potts and clock models, we re-express these in the basis (\ref{Eq_PottsBasis}), where the Hamiltonian has the general form:
\ba  \label{Eq_LInts}
\langle l'_1 l'_2 |\delta H_e | l_1 l_2 \rangle  &=&  \delta_{l'_1, l_1} \delta_{l_2', l_2}  \sum_{n=1}^{Q-1} \epsilon^{(e)}_n \cos \frac{ 2 \pi n (l_1 -l_2)}{Q} \n
\langle l' |\delta H_P | l \rangle  &=& \sum_{n=1}^{Q-1} \epsilon^{(P)}_n e^{ 2 \pi i (l'-l) \frac{n}{Q} }
\ea
The $Q$-state clock model is obtained for $\epsilon^{(e,P)}_{n} = \epsilon^{(e,P)}_{Q-n} = \delta_{n1}$; the Potts model occurs for $\epsilon_n^{(e)} \equiv 1, \epsilon_0^{(P)} =1, \epsilon_n^{(P)} = -1 (n>0)$, as described above.  More generally, we may consider any nearest neighbor $\mathbb{Z}_Q$ spin interactions.

Tuning the magnetic field leads to an even wider range of possibilities. If $\epsilon^{(P)}_n = \epsilon^{(P)}_{Q-n}$ then the field is again a superposition of clock-model type fields in different directions.  If $\epsilon^{(P)}_n \neq \epsilon^{(P)}_{Q-n}$ the model is chiral, in the sense that the transverse field has either a left- or a right- rotating eigenstate (but not both).   This can produce, for example, a special case of the chiral Potts model first described by Ref. \onlinecite{ChiralPotts}.  While we do not expect the choice of transverse field to affect the symmetry-broken phases, it could have an impact on the loci and nature of the phase transitions, which we will not attempt to describe.
We note, however, that there is no such freedom in the Ising or $Q=3$ Potts case -- and that for $Q>2$ the Potts transition is in any case first order in $3$D.

We therefore focus on the various possible interaction terms, and briefly describe the different possible phases allowed by symmetry-breaking in the spin model.  There are two possibilities: if $Q$ is prime, then condensing $\Phi^n$ for any $n$ will result in $\langle \Phi^k \rangle \neq 0$ for all $k$.  In this case all ferromagnetic symmetry-broken ground states in the spin model are related by a global $\mathbb{Z}_Q$ rotation.  The Potts Hamiltonian (\ref{Eq_HFull}) is special since it represents the unique choice of transverse field term and ferromagnetic coupling corresponding to a Hamiltonian that is exactly solvable in the limit $\alpha_T =0, \alpha_N >0$.  However, there are many choices of Hamiltonian which all undergo transitions breaking the same $\mathbb{Z}_Q$ symmetry.  Thus even for vanishing transverse field, there is a large family of parameters $\epsilon^{(e)}_n$ for which the Hamiltonian is in the same phase but not exactly solvable.  For $Q>3$ the clock model is one such Hamiltonian.

If $Q$ is not prime, then there will be some choices of $k$ for which we can have both $\langle \Phi^k \rangle \neq 0$ and  $\langle \Phi \rangle = 0$, giving two or more distinct phases in which different subgroups of $\mathbb{Z}_Q$ have been broken.
In the Potts basis of Eq.~(\ref{Eq_LInts}), this means that
\be
\langle l \rangle = v_0 \sum_{j=1}^{Q/k} \left( e^{\frac{ 2\pi i k}{Q} } \right)^{jl}
\ee
where we have taken the simplest case $\langle \Phi^{jk} \rangle \equiv v_0$ for all $j=1 \ldots Q/k$.  Because $Q/k$ is an integer, this results in a configuration of expectation values which is invariant under shifts in $l$ by a subgroup of $\mathbb{Z}_Q$.

To illustrate this possibility, we take $Q =4$.  We may individually set the values of the transverse field and ferromagnetic interaction for the two $\mathbb{Z}_4$ fields $\Phi, \Phi^3$ (whose  couplings must be the same for the Hamiltonian to be hermitian), and the $\mathbb{Z}_2$ field $\Phi^2$.  The ferromagnetic coupling of the spin Hamiltonian is
\be  \label{Eq_HAsht0}
H = J_2 \sum_{<ij>}  (\Phi_i \Phi^3_j + \Phi^3_i \Phi_j) + J_4 \sum_{<ij>}  \Phi^2_i \Phi^2_j
\ee
(Note that we are considering $\Phi^3$, $\Phi^2$ and $\Phi$ to be three independent fields here).  Taking
\ba
\frac{e^{i \pi/4} }{\sqrt{2} } ( S + i  \sigma ) & \Rightarrow & \Phi \ \ \ \ \ \
 -  \sigma S  \Rightarrow \Phi^2 \n
\frac{e^{-i \pi/4}}{\sqrt{2} } (S- i  \sigma) & \Rightarrow &\Phi^3
\ea
we see that Eq.~(\ref{Eq_HAsht0}) is equivalent to  the Ashkin-Teller model:
\be  \label{Eq_HAshT}
H = J_2 \sum  (S_i S_j + \sigma_i \sigma_j) + J_4 \sum  \sigma_i \sigma_j S_i S_j  \ \ \ .
\ee

If we choose the transverse field term
\be
H' = B_1 \left( S_{i}^{( x)} + \sigma_{i}^{(x)} \right) + B_2  S_{i}^{( x)} \sigma_i^{(x)}
\ee
then the quantum problem in non-vanishing transverse field is equivalent to the $3$D Ashkin-Teller model.

The phase portrait of the classical $3$D Ashkin-Teller model has been studied, for example, in Ref. \onlinecite{AshT}.
If all couplings are ferromagnetic, there are two distinct phases:  for $J_2 \neq 0$ the full $Z_4$ symmetry is broken, whereas If $J_2 =0$, then there is a ``$\Phi^2$" phase with $\langle \Phi^2 \rangle \neq 0$ but $\langle \Phi \rangle = 0$.    There are two Ising-like second order phase transitions separating the $\Phi^2$ phase from both the paramagnet and the fully symmetry-broken phase.  The two phase boundaries end at a tricritical point, after which the phase boundary separating the paramagnet from the fully broken phase is first order.

In summary, there are a wide range of Hamiltonians which will ultimately condense the same simple current in the topological lattice model, corresponding to different choices of $\epsilon_n^{(P,e)}$ in Eq.~(\ref{Eq_HPert1}).  Equally, if $Q$ is not prime we may condense simple currents of order $Q' = Q/n$, leading to new distinct symmetry-broken phases.

\subsection{Perturbations outside of the Potts model subspace}

Perturbations of the form (\ref{Eq_HPert1}) are special in that they do not introduce any excitations other than $\Phi^{(n)}$ vortices into the ground states of the system.  In this case the task of understanding the phase transition reduces to one of understanding a spin model.  This statement is, at second glance, a rather surprising one: we have reduced a question about topological orders, where the long-ranged statistical interactions mediated by gauge fields dictate the inter-particle interactions, to one about a spin model in which there are no gauge fields at all.  Essentially this is because the magnetic particles we condense behave like vortices in an abelian gauge theory, so that we may exploit an electric-magnetic duality to map the theory of vortices on the lattice onto a theory of `charges' (here $\mathbb{Z}_Q$ spins) on the dual lattice.

A generic perturbation to the Hamiltonian (\ref{Eq_HFull}) will introduce excitations outside of this pure $\Phi^n$ vortex sector into the ground state, however.
Here we consider the question of whether such perturbations qualitatively alter the phase diagram or critical behavior.  We present general arguments  that terms which weakly mix electric excitations with the ground states do not alter the critical behavior, although for $Q>2$ the pure Potts transition is in any case first order.  Perturbations generating other types of vortices correspond to annealed disorder in the (classical $3$D analogue of the) spin system, which we also expect to be irrelevant at the critical point.

We begin with a slightly perturbed version of our Hamiltonian:
\be
H = H_{LW} + H^{(1)}+
\epsilon H^{(E)}
\ee
Here,  $H_{LW}$ is the original Levin-Wen model, and $H^{(1)}$ includes the terms which tune the system through the condensation transition (for example, those given in Eqs. \ref{Eq_HFull}, or any perturbation discussed in Sect. \ref{sub:within}).  $H^{(E)}$ contains perturbations  which cannot be described in the effective spin model.
%create electric sources which become confined in the condensed phase, and we wish to know whether $H^{(E)}$ is a relevant perturbation at the critical point.
The possible perturbations fall into three classes: $H^{(E)}$ may create deconfined electric excitations, magnetic excitations (which are always deconfined), or confined electric excitations.  We consider each possibility in turn, to determine whether any of these are relevant perturbations at the critical point.

Sources of deconfined electric excitations have no effect on the $\mathbb{Z}_Q$ spin state; they act as the identity operator on the $\mathbb{Z}_Q$ spin model subspace.  Hence a perturbation which excites only such sources merely changes the short-distance characteristics of the topological ground states over which the transition occurs.  The altered ground states will still be compatible with our mapping to the spin model, however; hence the critical theory will be unaffected by this perturbation.

Perturbations exciting magnetic sources $v \neq \phi^n$ also fall into two categories, depending on whether $\phi^r \times v = v$ for some $r<Q$.  These map onto dynamical dilutions in the transverse field and in the sites of the spin model, respectively.  Specifically, suppose a plaquette $P$ is occupied by a generic vortex with flux $v$ in the uncondensed ground state.  Acting with the operator which creates $\Phi$ on this plaquette will produce a new vortex with flux $v' = \phi \times v$.  Since the product is unique, we may map this state onto a spin state by identifying $v'$ on this plaquette with the $\mathbb{Z}_Q$ spin normally associated with $\phi$.  The resultant mapping onto states of a spin model is qualitatively no different from the one used above, provided that $\phi^r \times v$ is distinct for each $r =0 ... Q-1$.  However, if $v \neq \phi^n$ then plaquettes carrying flux $v$ and $\phi \times v$ both have the same energy cost $\epsilon_m$, so that the effective Hamiltonian for the spin model now contains dynamical disorder, in the form of sites at which there is effectively no transverse field.  

When $\phi^r \times v =v$ for some $r$, the mapping to Potts spins cannot distinguish between pairs of $\mathbb{Z}_Q$ spins $x, x+r$ and is hence no longer one-to-one.  For example, in the $SU(2)_2\times SU(2)_2$ example of the previous section, plaquettes with flux $\frac{1}{2}$ are unaffected by the addition of a spin-$1$ vortex.  These thus effectively act like  (dynamic) dilutions in the lattice, meaning that on some sites the ferromagnetic Potts interaction is always satisfied and thus adds a constant to the overall energy, independent of the spin configuration.  (In the more general case these dilutions become sites where the NN interaction is satisfied provided that the $\mathbb{Z}_Q$ spins differ by any multiple of $r$).  This is reminiscent of the effect of annealed vacancies on the classical $3D$ Potts model, which     
has been studied in the context of the Blume-Emery-Griffiths\cite{BlumeEmeryGriffiths} model.  In this case, for small perturbations the vacancies are irrelevant at the Ising critical point\cite{Kimel}, but can drive the transition first order at larger vacancy concentrations.  We therefore expect this to be an irrelevant perturbation at the Ising fixed point, and that it will not alter the first-order character of the Potts transition. 
%As in $2D$\cite{Aizenman,CardyJacobsen,CardyJacobsenNPB},  the singularities associated with first-order Potts transitions are weakened by increasing the dilution concentration, driving the transition second-order at sufficiently high concentrations\cite{Chatelain,Ballesteros}.   For the Ising case ($Q=2$), where the transition is second-order in the absence of such disorder, such disorder is relevant at the critical point and leads to a different set of critical exponents\cite{Newman,Berche}.  

When electric sources for the confined edge labels are present in the ground state, the mapping to the $\mathbb{Z}_Q$ spin model necessarily breaks down.  These labels are associated with domain walls in the ferromagnetic Potts phase; including sources for these labels amounts in the Potts language to having open domain walls, which is impossible in the purely statistical mechanical picture.  To include these excitations we must include gauge fields in the dual theory, which account for the phase winding of branch cut singularities at the end of each domain wall.   In the presence of gauge fields the Landau framework no longer strictly applies as the order parameter is necessarily non-local.  

When $H^{(E)}$ contains source terms for confined excitations, therefore, we must explicitly consider whether these source terms are relevant at the critical point.
To do this, we consider the effect of such source terms on expectation values of local operators in perturbation theory.  Specifically, we may expand the ground-state wave-function in powers of $\epsilon$ according to:
\begin{widetext}
\ba \label{Eq_PsiPerts}
|\Psi \rangle  &=& |\Psi_0 \rangle + \sum_i  \frac{\epsilon}{ E_i -E_0 } \langle \Psi_i | H^{(E)} | \Psi_0 \rangle | \Psi_i \rangle  + \sum_{i, j}  \frac{\epsilon^2}{ (E_i -E_0)(E_j - E_0 ) } \langle \Psi_i | H^{(E)} | \Psi_j \rangle  \langle \Psi_j | H^{(E)} | \Psi_0 \rangle| \Psi_i \rangle
+ \ldots \ \ \ .
\ea
The expectation value of an operator $\hat{O}$ in the ground state (\ref{Eq_PsiPerts}) can be evaluated to a specified order in $\epsilon$:
\ba \label{Eq_OPerts}
\langle \Psi | \hat{O} | \Psi \rangle &=&\langle \Psi_0 | \hat{O} | \Psi_0 \rangle +\sum_i  \frac{\epsilon}{ E_i -E_0 }  \langle \Psi_0 | H^{(E)} | \Psi_i \rangle \langle \Psi_i  | \hat{O} | \Psi_0 \rangle  + h.c. \n
&& +\sum_i  \frac{\epsilon^2}{ ( E_i -E_0 ) (E_j - E_0) }  \langle \Psi_0 | H^{(E)} | \Psi_i \rangle \langle \Psi_i  | \hat{O} | \Psi_j \rangle  \langle \Psi_j | H^{(E)} | \Psi_0 \rangle  \n
&&
+  \sum_{i, j}  \frac{\epsilon^2}{ (E_i -E_0)(E_j - E_0 ) } \langle \Psi_i | H^{(E)} | \Psi_j \rangle  \langle \Psi_j | H^{(E)} | \Psi_0 \rangle  \langle \Psi_0  | \hat{O} | \Psi_i \rangle
\n
&&+ \ldots
\ea
\end{widetext}
Here we imagine working in a geometry where the ground-state is unique; we will return to the more general case presently.
The specific form of the higher-order terms will not be important for the qualitative arguments we present here; the germane point is that the order $\epsilon^n$ term in Eq.~(\ref{Eq_OPerts}) contains $n$ powers of the unperturbed electric source gap $\epsilon_V$ in the denominator, and $n$ powers of $H^{(E)}$ in the numerator sandwiched between various excited states.

Since $H^{(E)}$ creates open electric strings which do not exist in the unperturbed ground state, inner products involving the unperturbed ground state are non-vanishing only when the net effect of these applications of $H^{(E)}$ is to create some number of sources, move them some distance along the lattice, and re-anhiliate them.  Thus all of the non-vanishing terms in (\ref{Eq_OPerts}) can be expressed in terms of closed Wilson loop operators $W^{i}_C$ for a confined source $i$ along some curve $C$ in the lattice:
\be
\langle \Psi | \hat{O} | \Psi \rangle =\sum_{ \{ i_1, ... i_n \} }  \sum_{ \{ C_1, ... C_n \} } \alpha_{C_1, ... C_n}^{ i_1 ... i_n } \langle \Psi_0| W^{i_1}_{C_1} ... W^{i_n}_{C_n} | \Psi_0 \rangle
\ee
where $ \alpha_{C_1, ... C_n}^{ i_1 ... i_n } $ are coefficients which must be determined by the perturbation theory.  Here $n$ is determined by the order in perturbation theory to which the result will be computed.  A term where the total length of all the Wilson lines is length $m$ can only occur at order $m$ or higher in perturbation theory.

Since the Wilson loop operators $W^{i}_C$ can be mapped exactly onto operators in the $\mathbb{Z}_Q$ spin model, we may study the fate of the critical theory by asking what effect these have on the expectation of the spin model's Landau
free energy.  This amounts to asking whether the Wilson loop maps to a relevant or irrelevant operator near the critical point.

\begin{center}
\begin{figure}[h]
\subfigure[]{
\includegraphics[width=2.8in]{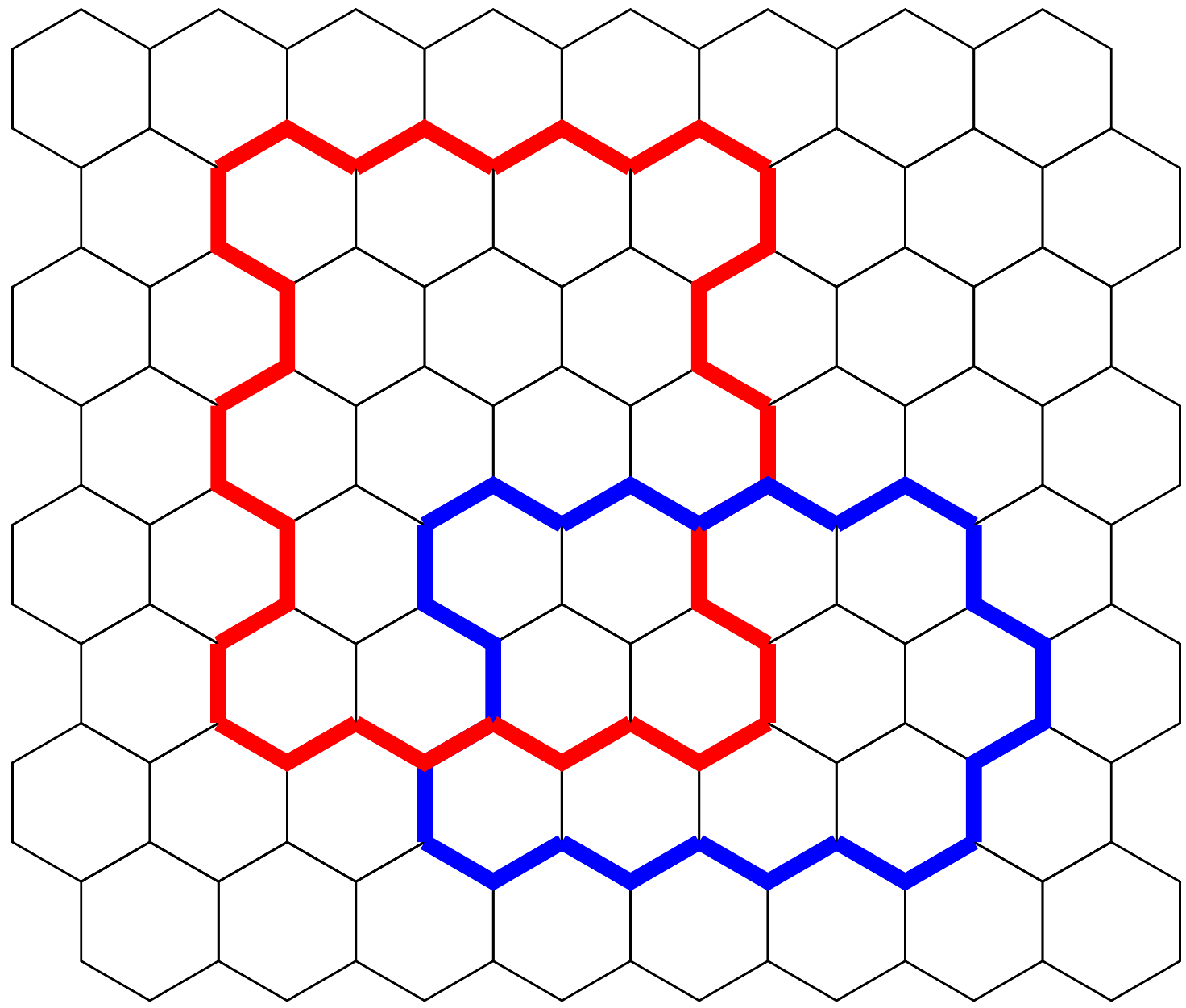}
}
\subfigure[]{ \includegraphics[width=2.8in]{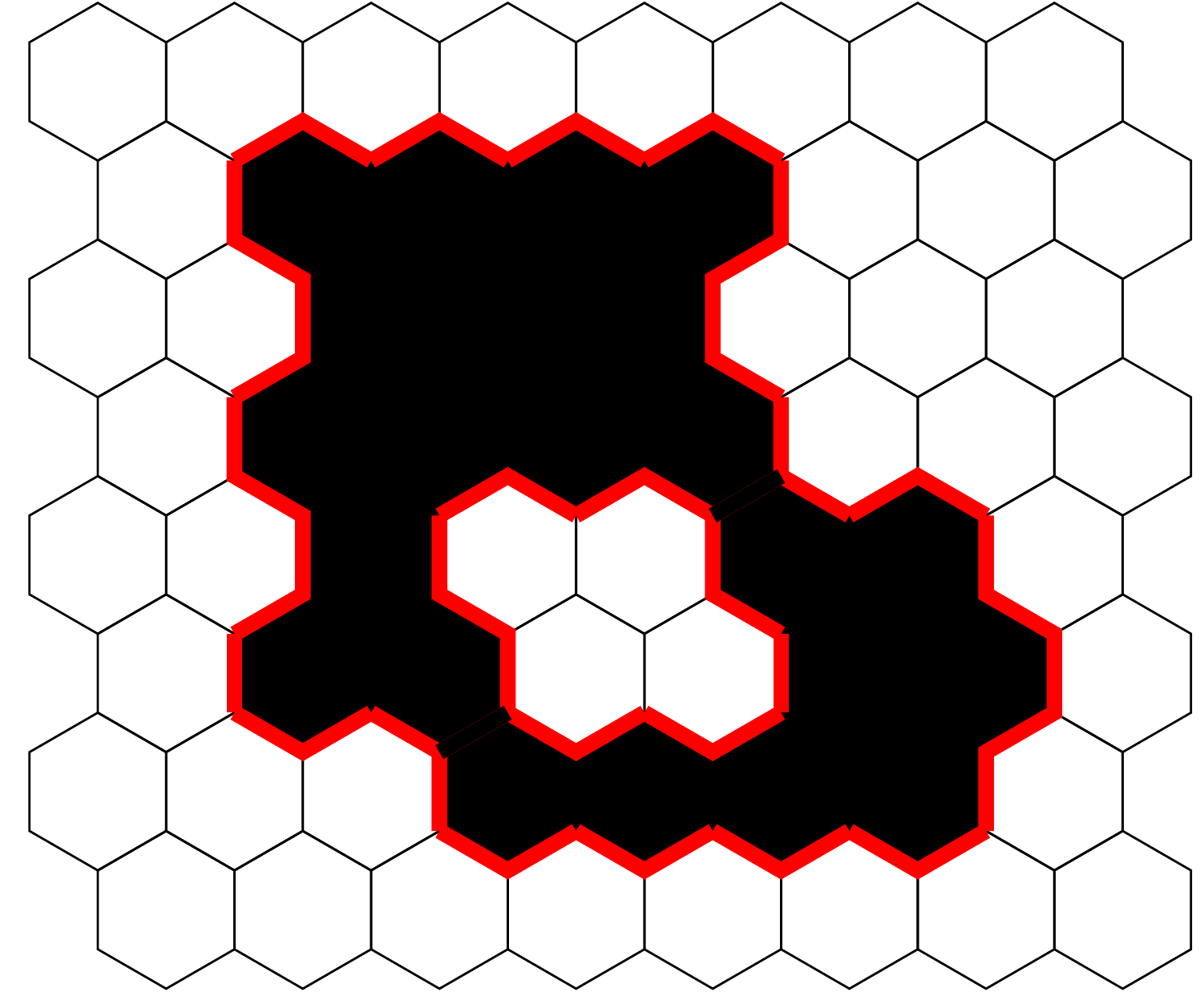}
}
\caption{ \label{Fig_WilsonLines} Mapping intersecting Wilson line operators to $\mathbb{Z}_Q$ domain walls.  (a) We specify a set of Wilson lines by their contour on the lattice and their linking.  Here the red line crosses once over and once under the blue line, so that the Wilson lines are linked.  (b) Up to constants, which are scale independent and determined by the topological order, the effect of a pair of linked Wilson lines is to flip all $\mathbb{Z}_Q$ spins encircled by each Wilson line by an appropriate amount.  (Here we show this for $Q=2$, where there is only one type of domain wall).  Spins encircled by two domain walls will be flipped twice.  }
\end{figure}
\end{center}

To answer this question, we must first consider in detail the form of the operators in the $\mathbb{Z}_Q$ spin description.  We will assume that $H^{(E)}$ contains only local terms, which either pair-create or move sources within some fixed radius on the lattice.
This implies that longer Wilson lines are suppressed by higher powers of $\epsilon/\epsilon_V$.  A single non self-intersecting Wilson line $W_C^i$ maps to an operator that flips a set of $\mathbb{Z}_Q$ spins to  generate a domain wall in which the $\mathbb{Z}_Q$ spin changes  by $i$ across the curve $C$ on the dual lattice.  Likewise multiple non-intersecting Wilson lines
map to an operator creating multiple domains.  To deal with intersecting Wilson lines, we must account not only for the positions of the domain walls, but also in general for an extra phase that occurs in the gauge theory when two Wilson lines cross.  This phase is dictated by the topological properties of the Wilson lines, and depends on the ordering of the Wilson line operators, as illustrated in Fig. \ref{Fig_WilsonLines}.   If $H^{(E)}$ creates only a small density of defects, then such crossings can occur only at relatively high orders in perturbation theory -- and are consequently suppressed by a high power of $\epsilon/ \epsilon_V$.

In Eq.~(\ref{Eq_PsiPerts}) we used non-degenerate perturbation theory, ignoring the topological ground-state degeneracy.  This is an appropriate starting point in general since the ground states in any case do not mix at low orders in perturbation theory.
For any finite-sized system, however, there will be some order in perturbation theory at which it becomes possible to create a Wilson line which wraps around one of the non-contractible curves (where these exist).  For confined sources this maps the $\mathbb{Z}_Q$ spin model into a sector with twisted boundary conditions; for deconfined sources it simply introduces a different ground state sector into the problem.   Here we assume that for local operators such effects occur at sufficiently high order in $\epsilon / \epsilon_V$ that they do not play an important role in the physics; indeed if they did, the ground state degeneracy of the topological phase would not be robust to their presence, indicating that the perturbation has altered the underlying topological order before $H^{(1)}$ tunes the system through the critical point of interest to us.

Restricting ourselves to orders in perturbation theory at which such operators cannot occur, we may assess the importance of $H^{(E)}$ at the critical point by considering whether the operator flipping clusters of spins is relevant.  When $H^{(E)}$ is local, at finite order in perturbation theory the maximum size of the clusters to be flipped is finite, and hence the operator is local at sufficiently long wavelengths, so that the Landau paradigm applies.
In the Ising model, where the phase transition is second order,
the Wilson line operators which flip clusters of spins are irrelevant at the critical point, and hence generically we expect that $H^{(E)}$ is as well.  This is in agreement with the results of Fradkin and Shenker\cite{FradkinShenker}, and has been verified numerically for the Ising transition between the Toric code and the vacuum by Ref. \onlinecite{VidalToric}, who established that electrical sources are irrelevant to the critical theory up to the point where the $\mathbb{Z}_2$ topological order itself is destroyed.

Thus  the fate of the phase diagram described in the previous section in the presence of generic perturbations is as follows.  The topological order of the gapped phases is unchanged so long as the perturbation does not close the excitation gap.  If the phase transition is second order, we may evaluate the effect of perturbations on the critical theory by leveraging the Landau-Ginzburg theory of the corresponding phase transition in the spin model, and considering whether the dual perturbation is relevant at the critical fixed point.   A small density of dynamic electric sources, which is dual to flipping clusters of spins, gives an irrelevant operator in the $\mathbb{Z}_Q$ spin description.  A small density of dynamic magnetic sources is dual to annealed disorder in the corresponding $3$D classical spin model, and is also not expected to alter the nature of the phase transition.

\section{Topological order of the condensed phase}  \label{FPhaseSect}

In Sect.  \ref{TransSect} we described a lattice Hamiltonian which we could tune exactly through a phase transition in which a magnetic bosonic excitation condensed.  We showed that the long-wavelength behavior near the phase transition could be mapped onto that of a ferromagnetic $\mathbb{Z}_Q$ spin model, and that in the condensed phase certain string labels -- corresponding to domain walls in the spin description -- became confined.  In fact, we found that there is a special point in the confined phase at which the Hamiltonian (\ref{Eq_HFull}) again becomes exactly solvable, as we may consistently project out the edge labels which become confined.
Here we study in more detail the physics of the confined phase, which can be understood by studying this second solvable point in the phase diagram.
Our objectives in doing this are two-fold: first, we will see how the features of the final-state spectrum (identified by Ref. \onlinecite{TSB}) arise in the lattice model, and discuss the explicit form of the final-state quasi-particle operators.  Second, we comment on the general structure of the lattice models that can be obtained as  condensates of doubled Chern-Simons theories.

To understand the physics of the condensed phase, it is useful to consider the topological properties of an $s$-wave superconductor\cite{SondhiSC}.  Before the onset of superconductivity (in the uncondensed phase), the system is well described in terms of electrons and holes (or Fermi liquid quasi-particles with charge $\pm e$), and the electromagnetic gauge field.  In the superconducting phase, the low-energy degrees of freedom are the Bogoliubov-deGennes quasi-particles and the superconducting vortex of flux $\frac{\hbar}{2e}$.
Electrons and holes of the original Fermi liquid theory are indistinguishable in the superconductor, since they are mixed by the condensate.   Further, the Meissner effect confines any gauge field flux unless it is appropriately quantized in units of $\frac{\hbar}{2e}$.  This is necessary so that the condensate is single-valued as it winds around the vortex; smaller flux quanta would necessarily result in a costly branch cut in the condensate wave-function.

Returning to the question of more general topological symmetry breaking transitions, a general prescription for obtaining the topological properties of the condensed phase is given by Bais and Slingerland\cite{TSB}, who identify three effects of condensation on the excitations of the original model.  These consist of confinement of excitations which braid non-trivially with the condensate (the Meissner effect), identification of pairs of excitations which are mixed by their interactions with the condensate (analogous to mixing of electrons and holes via scattering from Cooper pairs), and possible splitting of some excitations into multiple distinct quasi-particle types.
Here we will describe how these effects arise in the condensed phase of the lattice model, explaining how the spectrum predicted by Ref. \onlinecite{TSB} arises in practice from the excitations of the initial Levin-Wen model.

 \subsection{String operators and excitations }

 \begin{center}
 \begin{figure}[h]
 \subfigure[]{
 \includegraphics[width=2.8in]{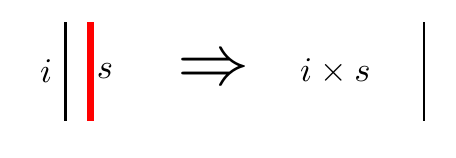} }
 \subfigure[]{
 \includegraphics[width=2.8in]{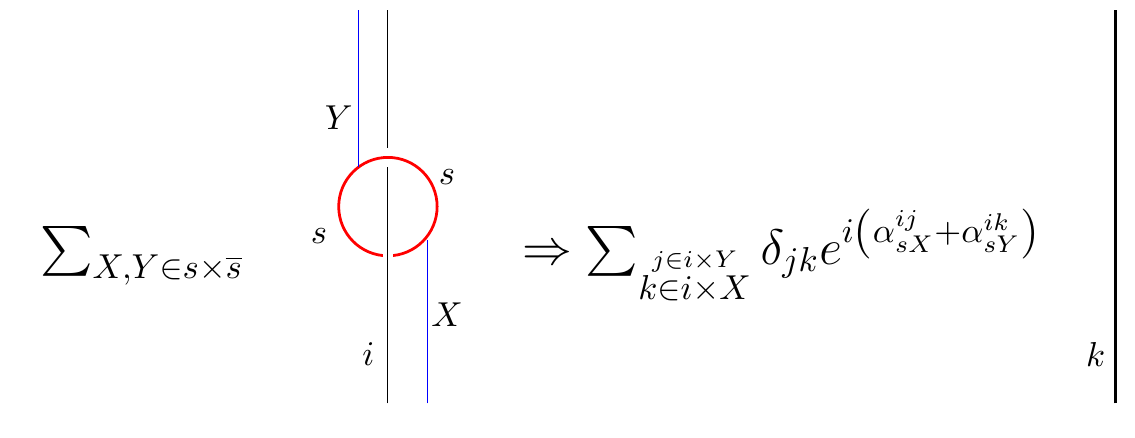} }
 \caption{
  \label{Fig_LWQP1} Quasi-particle creation operators in the Levin-Wen model are `string' operators which act on states so as to create excitations at each of the string's endpoints.  (a)  These strings can carry electric flux, in which case the string operator raises or lowers the label on each edge it runs parallel to, as shown.    In this figure a string operator with label $s$ acts on an edge carrying flux $i$ to form an edge with flux $i \times s$.
 (b) Strings can also carry magnetic flux, in which case the operator assigns a label-dependent phase to the wave function each time the string crosses an edge separating two plaquettes.  In the doubled Chern-Simons theory that we begin with, all quasi-particle strings are composed of composites of right- (electric only) and left- (composite electric and magnetic) components.  In this figure a magnetic string operator labeled $s \times \overline{s} $ acts on an edge $i$.
 The net electric flux is thus the combined flux from the two $s$ operators, which in general may take on multiple values.  Thus   the vortex string operator is the sum  of the string operator $\sum_{X, Y \in s \times \overline{s} }$.   For each $X, Y$ in the superposition the operator acts on the edge by raising its flux by $X$, and incurring a phase, as shown on the right.  This operation comes with an overall numerical pre-factor not shown here.  The chief importance of this pre-factor is to ensure that only diagrams in which $X$ and $Y$ raise the label $i$ by the same amount occur in the superposition.}
 \end{figure}
 \end{center}

 Before studying the excitations in the condensed phase, we must understand in more detail excitations in the uncondensed model.  These are most simply described in terms of the quasi-particles of the solvable Levin-Wen Hamiltonian, which we will describe briefly here.

As mentioned in the previous section, we may loosely speaking divide the excitations of the Hamiltonian (\ref{Eq_HLW}) into constituents which carry electric charge, and those which carry magnetic flux.
Operators creating electric charges on vertices $V_i$ and $V_j$ are open Wilson lines running from $V_i$ to $V_j$ (Figure \ref{Fig_LWQP1}).  For every edge label $i$ there is such a Wilson line, and consequently a distinct electric source.   As one would expect from lattice Yang-Mills theory, the Wilson line of the source $i$ raises the value of the electric flux on the edges that it traverses by $i$.

Magnetic sources -- such as the $\Phi = \phi_R \times \phi_L$ particle described above -- are created by an operator that essentially assigns a phase to each configuration of edge labels.   There is one type of magnetic source for each allowed edge label $i$.  Unlike in Yang-Mills theory, however, in the doubled Chern-Simons theory the elementary magnetic source ($i_L$) also carries electric charge\cite{CH}.  What we refer to as a magnetic $i$ vortex is the achiral particle $i_R \times i_L$.
 %The right handed particle $i_R$ involves an electric flux on each edge.  The left handed particle $i_L$, involves both a magnetic flux $i$ through a plaquette and an electric flux $\bar i$ along the edge (where $\bar i$ is the antiparticle of $i$ in the chiral theory).
When $i$ is a simple current, this is an excitation which violates only the plaquette terms of the Hamiltonian, exactly as we expect for a vortex.  When $i$ is not a simple current, however, its action on an edge $e$ involves  a superposition of terms, each of which raises $e$ by an element of $i \times \overline{i}$.
The precise form of this edge operator for a magnetic source is described in detail in Appendix \ref{QPApp} and in Ref.~\onlinecite{CH}.

\subsection{Confinement and identification in the condensed phase}

We begin by understanding how the first two phenomena -- identification of some excitations, and confinement of others -- occur in the condensed phase of (\ref{Eq_HFull}).  Essentially, this is no different than in the superconductor.  First, as we have already seen, excitations which braid non-trivially with the condensate (those for which $q_i \neq 0$) are confined in the condensed phase.  This is analogous to the Meissner effect, since their braiding statistics with the condensate ensure that they engender branch cuts (or domain walls, in the $\mathbb{Z}_Q$ spin picture) in the condensate wave function.
Further, after condensation, pairs of particles which are mixed by scattering with the condensate are indistinguishable (or identified) in the final topological theory, much like electrons and holes in a superconductor.

We have already demonstrated confinement explicitly in the previous section.  Specifically, the arguments used to show that closed domain walls incur a linear energy cost apply equally well to static sources: though their energy can exceed the cost $2 \epsilon_V$ of spontaneously creating a new pair of sources, the number of these sources is conserved by the Hamiltonian (\ref{Eq_HFull}) and such processes cannot occur.  At the solvable point where edge labels with $q_i \neq 0$ do not appear in the ground state, the energetic cost of adding a pair of sources with $q_i \neq 0$  is linear in the length of the string separating them, as there must be a string of edges with confined labels which connects the pair.  Away from the solvable point there is in general a small admixture of closed strings of the confined $q_i \neq 0 $ labels in the ground state, but as this is relatively small (domain walls are confined) the energetic cost of a pair of confined sources will still scale linearly with their separation.  (As in QCD, perturbations mixing confined {\it sources} (or open confined strings) with the ground state lead to difficulties in identifying the confinement potential, as the string joining a pair of test sources can be broken into shorter segments by creating confined particle-anti-particle pairs).

To show in detail how identification occurs, we note that in the ground state of Eq.~\ref{Eq_HFull} at $\alpha_T =0, \alpha_N =1$ each plaquette is in a superposition of states with excitations $Id, \Phi, .. \Phi^{Q-1}$.  Hence any string operator $\hat{W}_{\Phi^n \times s}$ applied to this phase will result in the same superposition of sources with the labels $s, \Phi \times s, .. \Phi^{Q-1} \times s$. (Here $s$ is a particle of the doubled theory).  In the condensed phase, therefore, there is a single string operator associated with the sources
\be
s, \Phi \times s, ... \Phi^{Q-1} \times s
\ee
and correspondingly all of these should be identified as creation operators for the same quasi-particle.

The rules for mixing here are inherited from the `fusion' rules, which specify how to combine representation labels on the edges of the lattice.  In a topological theory there are rather stringent consistency constraints on these rules (see, for example, Ref. \onlinecite{BondersonThesis}), but for the present discussion the important point is that the excitations naturally mix with the condensate in a specific way.
When $\Phi$ is a simple current, the fusion rules  specify that $\Phi \times s$ is a {\it single} quasi-particle species.

We illustrate the effect of confinement and identification on the spectrum with a few examples.  It is convenient to define the notational convention
\be
(r, s) \equiv r_R \times s_L
\ee
for excitations in the doubled Chern-simons theory.  In this notation, the magnetic $\Phi = \phi_R \times \phi_L$ particle that we condense has the form $(\phi, \phi)$.

\subsubsection{ $SU(2)_k$ for $k$ odd }

The labels appropriate to $SU(2)_k$ Chern-Simons theory are the total spin of representations $s =0 (\equiv Id), 1/2, ... k/2$.  We condense the $(k/2, k/2)$ excitation, which is an achiral simple current of order $2$ (i.e., a $\mathbb{Z}_2$ boson).

To find the confined particles, we note that the $S$-matrix elements are
$\frac{1}{ \Delta_s}  S_{\frac{k}{2}, s} = (-1)^{2s}$ (see discussion near Eq.~\ref{Eq_Smatrix}).  Hence edge labels with half-integer spin are confined (they all are effectively mapped to domain walls of the Ising model), while those with integer spin remain deconfined.  In the solvable limit, this implies that the plaquette projector also contains only integer spin labels.  Since the edge label (or electric flux) depends on the combined electric fluxes from the right- and left-handed components of each excitation, the deconfined particle types are:
\be
(r,s) \ \ \ \mbox{ and } \ \ \ \left (\frac{r}{2}, \frac{s}{2}  \right )
\ee
for $r$ and $s$ integer.

Fusing the $(k/2, k/2)$ particle with other excitations in the theory gives:
\be \label{Eq_SU2Id}
(k/2, k/2) \times (r,s) = (k/2-r, k/2 -s) \ \ \ .
\ee
As promised, the rules for adding angular momenta in the Chern-Simons theory have been deformed such that the product on the left-hand side gives a unique result on the right.
 Eq.~(\ref{Eq_SU2Id}) therefore identifies pairs of excitations
\be
(r,s) \equiv \left( \frac{ k-2r}{2}, \frac{ k-2s}{2} \right )
\ee
When $k$ is odd, if $r$ and $s$ are integers, then both labels on the right are half-integer.  Thus we may eliminate the half-integer labels from the theory completely, and identify each excitation by the appropriate integer labels.

In this case, this is the whole story: condensation has merely eliminated all half-integer spin labels from the theory.  The result is a doubled $SO(3)_k$ Chern-Simons theory\footnote{By this, we mean the integer spin sector of $SU(2)_k$.  }.

\subsubsection{ $SU(2)_k$ for $k$ even }

After condensing the $(k/2,k/2)$ boson, if $k$ is even, our result will differ from that described above in that $k/2$ is an integer, so that fusion with the condensate identifies integer pairs with integer pairs, and half-integer pairs with half-integer pairs.  Hence we can no longer eliminate the half-integer labels entirely from the theory, as they may still appear in the guise of excitations of the form $\left( r/2, s/2 \right )$ with $r$ and $s$ odd, even though they will no longer enter as deconfined edge labels.
This suggests that the topological description of the condensed phase is not a doubled Chern-Simons theory as we shall see further below.

A second peculiarity of the identifications is that the quasi-particle $(k/4, k/4)$ maps to itself under fusion with the condensed $(k/2, k/2)$.  We will see that this is not unrelated to the fact that the quasi-particle spectrum cannot be simply separated into right- and left- handed components when we discuss splitting.

\subsection{Splitting}
\label{sub:split1}

The final possible effect of condensation is that some excitations will split into multiple distinct  quasi-particle species.  (This phenomenon occurs in spontaneously broken non-Abelian gauge theories, but not in the superconducting example discussed above).
It is important to note that this splitting does not change the total dimension of the low-energy Hilbert space.  Instead, it splits a $2$- (or more generally $d$-) dimensional internal Hilbert space of the excitation in the uncondensed phase into multiple $1$- (or $d/n$) -dimensional Hilbert spaces.

An intuitive understanding of this splitting can be gained by considering how it arises in non-Abelian gauge theories.  Essentially what happens is that the excitation in question transforms under a $d$-dimensional representation of an internal symmetry which is broken in the condensed phase into multiple lower-dimensional representations of the residual symmetry group.  For example, in a theory with unbroken $SU(2)$ symmetry, a spin-$1/2$ particle is associated with a $2$-dimensional internal Hilbert space.  (In other words, we may express particles transforming in the fundamental representation of $SU(2)$ as $2$-component vectors).  If we break the $SU(2)$ symmetry by condensing a spin-$1$- Higgs field, this $2$-component vector can be separated into its spin-up and spin-down constituents, which are now no longer related by symmetry.  The residual $U(1)$ gauge transformation acts on these as:
\be
c^\dag_\uparrow \rightarrow e^{i \theta} c^\dag_\uparrow \ \ \  c^\dag_\downarrow \rightarrow e^{-i \theta} c^\dag_\downarrow
\ee
In other words, the $2$-dimensional representation of $SU(2)$ has separated into two $1$-dimensional representations of $U(1)$ (here carrying opposite charges).  The total dimension of the Hilbert space associated with each particle creation operator $c^\dag$ is still $2$, though we now have distinguishable spin-up and spin-down excitations.

One indicator that such a splitting had to happen in the above example can be found in the rules for combining the representations of $SU(2)$.
% To be specific, let us assume that the spin-$1$ particle that condenses is the triplet bound state of two spin-$1/2$ particles.
Combining two spin $1/2$ excitations gives:
\be \label{Eq_HalfAdd}
\frac{1}{2} \times \frac{1}{2} = 0 + 1
\ee
which is a superposition of the (gauge-neutral) singlet and the spin-$1$ triplet excitations.  Before condensation, these are distinct excitations, as they transform in different representations of the symmetry group.  After condensation, however, the residual $U(1)$ symmetry group {\it cannot} distinguish between the singlet and triplet states.
%, since the (spin-$1$) condensate is not charged under the residual symmetry group.  
Instead, it is sensitive only to the $l_z$ eigenvalue of each state, rather than to $l$.  Labelling states on the right-hand side of  (\ref{Eq_HalfAdd}) by their $l_z$ eigenvalues, we have:
\be
\frac{1}{2} \times \frac{1}{2} = 0 + 0 + 1 + (-1)
\ee
which is to say, on the right-hand side we obtain two distinct copies of the singlet ($l_z =0$) representation of $U(1)$.  (In this case, these are $c^\dag_{i \uparrow} c^\dag_{j \downarrow} \pm  c^\dag_{j \uparrow} c^\dag_{i \downarrow}$).  The rules of representation theory dictate that two copies of the singlet can be obtained on the right-hand side only if there are two distinct excitations on the left (which we may choose to be $c^\dag_{\uparrow}, c^\dag_{\downarrow}$).  Hence we conclude, solely by examining the rules for addition of angular momenta, that the spin-$1/2$ excitation {\it had} to split into two distinct $1$-dimensional representations (and hence $2$ distinct quasi-particle types) in the condensed phase.

This same logic about combining representations in the condensed phase applies to the more general framework of topological symmetry breaking\cite{TSB,TSBPRL,TSBPRL2}.  This will  indicate that, in some of the examples of the type discussed here, certain excitations split into multiple distinct quasi-particles after condensation.  We will first review the criteria for splitting to occur, and then explore how it arises in practice in the lattice models.

\subsubsection{Determining whether quasi-particles split}

The generalization of the criteria we found above in $SU(2)$ Yang-Mills theory for splitting after condensation is as follows.  Suppose $\Phi$ is a simple current (with $\Phi^Q =1$) which we will condense.  Then an excitation $r$ will be split in the condensed phase if
\be
\Phi^k \times r = r
\ee
for $k < Q$.  As in the example above, the reason for the splitting is that in the condensed phase, $r \times \overline{r}$ contains multiple copies of the trivial representation, and hence must be split into multiple particle types if the representation theory is to remain consistent.
Specifically, the representation $\overline{r}$ is by definition the one which combines with $r$ to give the the singlet representation (plus some other representations, in general).  Thus we have:
\ba
&\overline{r} \times r =& Id \, + ... \n
=&\overline{r} \times\left( r \times \Phi^k  \right) =& \Phi^k + ... \n
=&\overline{r} \times\left( r \times \Phi^Q   \right) =& \Phi^Q + ...
\ea
which implies that
\be  \label{Eq_Splits}
\overline{r} \times r  = Id + \Phi^k  + \Phi^{2k} + ... + \Phi^Q +  ... \ \ \ .
\ee
That is, the tensor product of representations $r$ and $\overline{r}$ contains (among other things) all powers of $\Phi^k$ (mod $Q$).  Since in the condensed phase any power of $\Phi$ is identified with the trivial representation (since, as before, the condensate is by definition in the singlet representation of the residual symmetry group), this gives $\frac{Q}{k}$ singlet representations on the right-hand side of  (\ref{Eq_Splits}).  Excitations in representation $r$ before condensation consequently split into $\frac{Q}{k}$ distinguishable particle types in the condensed phase.

It is a feature of the representation theory that the total dimension of the Hilbert space (total quantum dimension) is preserved by this splitting.  In particular, if $r$ is $1$-dimensional then it follows that $k=Q$ and the excitation cannot split.  This is rather obvious in the case that $r$ is truly a representation of a non-Abelian symmetry group; however, it also holds true in the truncated representation theory germane to the Chern-Simons lattice models considered here\cite{TSB}.

\subsubsection{Splitting on the lattice}

Armed with this simple criterion to understand when some excitations in the condensed phase will split, we now turn to the question of how this splitting manifests itself on the lattice.  In the example given above,
% in section \ref{sub:split1}
we could explicitly identify 
the two split spin-$1/2$ particles as we knew precisely the form of the residual symmetry generator's action on the members of the original $SU(2)$ multiplet.  In the lattice model (\ref{Eq_HFull}) we do not have access to this information.  However, we will be able to identify a set of labels (analogous to $S^z$ in the example above) which are indeterminate before condensation but separately conserved in the condensed phase.

Though this may seem like a rather trivial exercise in practice, since we have already argued on general grounds that such splittings {\it must} occur, it is actually important to demonstrate that splitting occurs in order to conclude that the condensed phase does indeed represent a consistent topological theory.  As we discuss briefly in the conclusions, there are situations where it is not clear that this is the case.

Let us begin with an example, and consider condensing the spin-$1$ excitation in doubled $SU(2)_2$ Chern-Simons theory.  The fusion rules for this theory are given in Eq.~(\ref{Eq_Fuse}), and in particular stipulate that
\be
(1,1) \times \left( \frac{1}{2}, \frac{1}{2} \right) = \left( \frac{1}{2}, \frac{1}{2} \right)
\ee
so that the achiral spin-$1/2$ particle must split into $2$ distinct excitations after condensation of the $(1,1)$ boson.  To identify these distinct excitations, we first note that the achiral spin-$1/2$ particle carries a magnetic flux from its $1/2_L$ component, and an electric flux from the combination of its $1/2_L$ and $1/2_R$ components.  Since $1/2 \times 1/2 = 0 +1$, the electric flux associated with this excitation can be either $0$ or $1$ on a particular edge.  The precise form of the string operator dictates that it may change between $0$ and $1$ when the string crosses between two plaquettes over an edge carrying a spin-$1/2$ label (see Appendix \ref{QPApp}).
At the solvable point in the condensed phase, however, the spin-$1/2$ edge labels have been completely eliminated from the Hilbert space.  Thus in this limit, a  $\left( \frac{1}{2}, \frac{1}{2} \right)$ particle is associated with an electric flux which is either $0$ or $1$ at all points along the string operator (and consequently, also at the vertices on which the string terminates).  We may therefore identify two distinct quasi-particle types, $\left( \frac{1}{2}, \frac{1}{2} \right)_0$ and $\left( \frac{1}{2}, \frac{1}{2} \right)_1$.

The key point here is that prior to condensation, a string which is purely of the $\left( \frac{1}{2}, \frac{1}{2} \right)_0$ or $\left( \frac{1}{2}, \frac{1}{2} \right)_1$ type is not topological.  That is, even at the exactly solvable point this string is physically observable -- whereas the string operators of Ref. \onlinecite{LW} create strings for which only the end-points have physical meaning.  The $\left( \frac{1}{2}, \frac{1}{2} \right)_0$ and $\left( \frac{1}{2}, \frac{1}{2} \right)_1$ strings, however, create a spin-$1/2$ vortex at each endpoint, together with a string of plaquettes in a superposition of the ground state and the spin $1$-vortex (the $(1,1)$ boson) excited along its trajectory, as explained in Appendix \ref{QPApp}.  Hence in the uncondensed phase, the eigenstate $\left( \frac{1}{2}, \frac{1}{2} \right)$ is associated with a $2$-dimensional Hilbert space (of electric flux $0$ or $1$).  In the condensed phase, where the string of possible spin $1$-vortex excitations is undetectable, $\left( \frac{1}{2}, \frac{1}{2} \right)_0$ and $\left( \frac{1}{2}, \frac{1}{2} \right)_1$ are two distinct topological quasi-particle creation operators.

This example, though relatively simple, illustrates precisely how splitting occurs in general condensates.
First, notice that if $\left( \phi^k, \phi^k \right) \times \left( r,s  \right) = \left( r,s  \right)$, then we must have both $\phi^k \times r =r$ and $\phi^k \times s = s$.  
The electric flux associated with this excitation is a superposition of 
\be
r \times s = \sum_{l\in r \times s} l  \ \ \ .
\ee
After condensation, we find that the sum on the right-hand side will split into subsets of labels which do not mix once the confined edge labels have been projected out of the theory.  It is possible to deduce from the fact that $\Phi^k \times (r,s) =(r,s)$, and that $\Phi$ acts as the identity in the condensed phase, that there are  $\frac{Q}{k}$ excitations which are topological in the condensed phase.  (As above, in the uncondensed phase there is only one).
A (rather technical) proof of this fact is given in Appendix \ref{SplittingApp}.

Thus we find that the excitations in the condensed phase split into $\frac{Q}{k}$ distinct species of excitation, as required for a consistent topological phase.

\subsection{Structure of the final theory: Examples} \label{ExampleSect}

Having established the nature of the spectrum in the condensed phase, it is worth pausing to take stock of the variety of possible topological phases which can be created in this way by condensing a magnetic simple current in a doubled Chern-Simons lattice model.  We will illustrate this with a series of examples.

\subsubsection{ Theories without splitting}

The simplest case we may consider is that of a theory in which no particles split in the condensed phase.  The properties of the spectrum here are determined by confinement and identification alone.  Depending on the nature of the condensed excitation, the final theory may be a new doubled Chern-Simons theory (with a gauge group that is a quotient group of the original, as generally occurs when vortices are condensed\cite{TSBPRL2}).  Alternatively, it may be a theory in which not all excitations can be decomposed into separate right- and left- handed chiral components.

The difference between these two cases is determined by $S_{\phi^k \phi} = e^{\frac{ 2 \pi i k q_\phi}{Q} }$, where we condense $(\phi, \phi)$.  If $q_{\phi^k} \neq 0$ for all $k<Q$, then in the condensed phase each set of identified quasi-particles either is confined or contains one element which is a composite of deconfined right- and left-handed labels.  Specifically, any deconfined excitation $(a, b)$ has $q_a = q_b$.  Since charge is additive under fusion with $\phi^k$, and since by assumption the charge of $\phi^k$ spans all possible values, there is some $r$ for which $q_{\phi^r} = - q_a$.  Hence $(\phi^r, \phi^r) \times  (a,b) $ is composed of two string-types which both have $q$-charge $0$.  Hence every excitation in these theories can be viewed as a composite  of two deconfined particle types. In this case, the condensed theory is just two opposite chirality copies of a subset of the particles of the original chiral Chern-Simons theory.

If $q_{\phi^k} = 0$ for some $k<Q$, however, the above result need not hold.  It is easy to construct examples of this in Abelian Chern-Simons theories.  For example, the $k$ particle in $U(1)_{2k} $ has:
\be
 k \times k \equiv Id \ \ \  \ \ \ S_{k j} = (-1)^j.
\ee
If $k$ is odd, then all deconfined excitations can be expressed in terms of pairs $(2j, 2l)$ of deconfined excitations, and the theory is again a tensor product $U(1)_k \times \overline{U(1)}_k$ of identical right- and left- handed Abelian Chern-Simons theories.  If $k$ is even, the deconfined excitations fall into two classes: $(2j, 2l)$ and $(2j +1, 2l+1)$ which are not equivalent under fusion with $(k, k)$.  (When $k$ is odd these odd and even sectors are identified).  Hence here the spectrum is not a direct product of two chiral components, as neither component of the odd excitations can exist in isolation.  % (In both cases the final result is topologically a $Z_k$ gauge theory with matter).

\subsubsection{$SU(2)_2$ }

The simplest case where splitting does occur is after condensation of the achiral spin $1$ excitation in a doubled $SU(2)_2$ Chern-Simons theory\cite{TSBShortP}.  As noted above, the $(1/2, 1/2)$ excitation splits into two components in the condensed phase, distinguished by their electric flux (which may be that of the spin singlet or spin triplet).  Since the chiral spin-$1/2$ excitations are confined, this leaves us with the following $3$ quasi-particles in the condensed phase:
\be
(0,1) \equiv (1,0) \ \ \ \ \ \  \left ( \frac{1}{2}, \frac{1}{2} \right)_{0}  \ \ \  \ \ \left ( \frac{1}{2}, \frac{1}{2} \right)_{1} \ \ \ .
\ee
The first of these, which is a purely electric source in our lattice model, is a fermion.  The other two (which we identify as purely magnetic, and both magnetic and electric) are bosons (but here with relative semionic statistics).

These excitations give precisely the spectrum of Kitaev's Toric code\cite{KitaevToric} (or $Z_2$ gauge theory with matter).  Indeed, at the solvable point in the condensed phase, where we eliminate all spin-$1/2$ edge labels from the theory, we may use the edge-labeling scheme:
\be
\sigma^x_e =\begin{cases} 1 \mbox{ if } \ \ i_e = 0 \\
-1 \mbox{ if } \ \ i_e = 1  \ \ \ .
\end{cases}
\ee
In this basis, and dropping the terms $- (-1)^{n_\sigma}$ (which we take always to be $-1$, since $n_\sigma \equiv 0$)
 the Hamiltonian is precisely that of the Toric code:
\be
H = -  \sum_V \prod_e \sigma^x_e -  \sum_P \prod_e \sigma^z_e
\ee

In matching the excitation spectra to that of the Toric code, we must contend with one subtlety of the condensed phase -- namely, the purely electric source is fermionic, rather than bosonic as it should be for the $Z_2$ gauge theory.
The reason is that operator which creates the $(1,0)$ excitation is an electric-type string which raises the spin on each edge by $1$ (mod 2); however, it also obtains a phase of $\sigma^x_e$ for each edge $e$ it crosses.  The operator creating $\left ( \frac{1}{2}, \frac{1}{2} \right)_{0}$ excitations is a magnetic-type string which assigns a phase of $-1$ for each edge of spin $1$ (mod $2$).   $\left ( \frac{1}{2}, \frac{1}{2} \right)_{1}$ is the operator which raises the spin by $1$ without inducing any phases -- and is mutually semionic relative to both $(1,0)$ and $\left ( \frac{1}{2}, \frac{1}{2} \right)_{0}$.   In the spin basis, this gives the quasi-particle operators the final form:
\ba
  \hat{s}_{(1/2, 1/2)_0} &=& \prod_{e} \sigma^x_{e}  \ \ \ \ \ \hat{s}_{(1/2, 1/2)_1} = \prod_{e} \sigma^z_{e}  \n
  \hat{s}_{(1,0)} &=&   \hat{s}_{(1/2, 1/2)_0}   \hat{s}_{(1/2, 1/2)_1}
\ea
 which identifies $\left ( \frac{1}{2}, \frac{1}{2} \right)_{1}$ as the electric source of the $Z_2$ gauge theory, $\left ( \frac{1}{2}, \frac{1}{2} \right)_{0}$ as the $Z_2$ magnetic source, and $(1,0)$ as their (fermionic) composite.

\subsubsection{$SU(2)_k$ for $k$ even}

We may generalize some of the features of the $SU(2)_2$ example above to $SU(2)_k$ for general even $k$, condensing the $(k/2,k/2)$ vortex.  The deconfined excitations here have net integer spin on each link, and hence must have the form $(i,j)$ with $i$ and $j$
either both integer or both half-integer spins.  This gives $ 2 \left( \frac{k}{2} \right)^2  +2 \frac{k}{2} +1$  excitations before identification.  Of these excitations, all but $\left( \frac{k}{4}, \frac{k}{4} \right)$ get identified in pairs; $\left( \frac{k}{4}, \frac{k}{4} \right)$ in fact splits into two excitations.  Thus we obtain a total of $\left(\frac{k}{2}\right)^2+ \frac{k}{2}+ 2$ excitations.  For $k>2$ the number of excitations is generally not a perfect square, so that the topological order cannot be that of a doubled Chern-Simons theory (or any double $T \times \overline{T}$ of a valid topological theory).

When $k=2$, we showed above that the condensed phase is  the topological limit of a discrete gauge theory.  For  $k=4$, the condensed phase is also described by a discrete gauge theory, in this case a twisted version of the non-abelian gauge group $D_3$.  For $k>4$, however, the topological order of the condensed phase cannot be that of a discrete gauge theory, since some of the deconfined particles have non-integer quantum dimensions.

We can nonetheless write the complete set of string operators for these theories.  There is an even sector of string operators which are composed entirely of the integer spins of the uncondensed phase, and an odd sector of string operators composed of pairs $i_L \times j_R$ of half-integer spins of the uncondensed theory.  The even sector consists of excitations which can be constructed using only strings which correspond to deconfined labels.  Specifically, we have:
\ba
\mbox{simple } : &\ \ \ \ \begin{cases}   i_L, i_R \ \ \ (i=Id... \frac{k}{2} -1 ) \n
 \frac{k}{2} \ \ \ \ \ \ \  (\mbox{non-chiral}) 
 \end{cases}\n
\mbox{ composite }: &
i_R\times j_L \ \ \  i\leq \frac{k}{4}, j< \frac{k}{2}
\ea
with $i$ and $j$ integer.
Because of the identifications (which identify even excitations with even, and odd with odd, unlike the scenario for odd $k$), the excitations in the even sector contain only some of the allowed composites which we can construct from the chiral string operators $i_{L,R}$.  In addition, there is only a single string operator $\frac{k}{2}_L \equiv \frac{k}{2}_R$ associated with the $\frac{k}{2}$ particle, whose chirality is no longer well-defined after condensation.  Indeed, any particle of the form $x_R \times \left ( \frac{k}{2}-x \right )_L$ is non-chiral in the sense that in the condensed phase it is indistinguishable from the opposite chirality excitation $x_L \times \left ( \frac{k}{2}-x \right )_R$.

In addition to these, there are excitations which cannot be composed of simple string operators.  That is, they are composites of pairs of half-integer spin string operators.  These composites have become irreducible in the condensed phase, where the individual half-integer spin strings have been confined.  These excitations are:
\be
i_R\times j_L \ \ \  i \leq \frac{k}{4}, j< \frac{k}{2}
\ee
with $i$ and $j$ half-integer.

The split $\left( \frac{k}{4}, \frac{k}{4} \right)$ particle is in the even sector for $k=0$ (mod $4$), and the odd sector for $k=2$ (mod $4$).

\subsubsection{Drinfeld Doubling}

The general structure of the construction here is as follows.   At the solvable point in the condensed phase, we may project onto states composed only of deconfined edge labels (those which braid trivially with the condensed magnetic excitation), from which our fixed-point  Hamiltonian for the condensed phase (Eq. \ref{Eq_HEffC}) and its low-energy excited states are constructed.  When $q_{\phi^k} \neq 0$ for any $k<Q$,  all excitations in the final theory can be constructed from string operators containing only these deconfined edge labels.
When $q_{\phi^k} = 0$ for some $k<Q$ (which must occur if there is splitting, but may occur in other examples as well, such as the Abelian theories discussed above), we generally find that not all excitations in the final model can be expressed in terms of strings corresponding to deconfined edge labels.  Nevertheless, the resulting topological theory is equivalent to a Levin-Wen model built on the category of deconfined edge labels only.   In general a Levin-Wen model built from a category produces a topological theory known as the Drinfeld double of the category.   In our case we identify the ``non-simple" string operators of Ref. \onlinecite{LW} as those which cannot be constructed from deconfined edge labels alone, but require additional phases to account for the fact that they are composites of pairs of confined electric sources.

The simple protocol outlined here shows how certain Drinfeld doubles can be interpreted physically as the outcome of condensation in a Chern-Simons theory.  Many of the examples given above are in any case discrete gauge theories, which we could alternatively understand as the result of Higgsing a continuous Maxwell or Yang-Mills theory.  However, some (for example, those obtained by condensing the $(k, k)$ particle in a doubled $SU(2)_{2 k}$ Chern-Simons theory with $k>2$) are not; this approach offers a physical mechanism for the origin of these states from models which can at least be understood in terms of continuous field theories.
In general, our approach suggests that we may view some Drinfeld doubles as arising because an achiral excitation condenses in a doubled Chern-Simons theory.  Specifically, because the condensate is achiral, time reversal symmetry is preserved on both sides of the phase transition, though the spectrum of the final theory cannot always be decomposed into decoupled right- and left- chiral sectors.

\section{Conclusions}

In this work we have given an explicit realization of topological symmetry breaking\cite{TSB} in lattice models.  By constructing a lattice Hamiltonian that can be tuned between two solvable Levin-Wen points, via a condensation transition, we may map the topological symmetry breaking transition explicitly onto a $2+1$D transverse-field Potts transition.  The phase transition can be understood by studying the dual Potts description, to which the Landau formalism applies if the transition is second order.  Though this duality is precisely valid only for a very special trajectory through the phase diagram, we argue perturbatively that the effect of small deviations from this trajectory can be understood within the Landau theory of the spin model,  and will be irrelevant at the critical point.  This gives a general framework to clarify the relationship between phase transitions separating different topological orders (such as those described by ref. \onlinecite{WenBarkeshliLong}) and phase transitions of the Landau type.  Further, we have studied the properties of the condensed phase, and identify the complete set of quasi-particle creation operators required for a consistent topological phase.

The type of transitions we discuss here are special in two ways.  First, in topological theories,  two bosons of the same type generally combine to give a variety of other species of bosons.  This is analogous to combining spins, where for example $1/2 \times 1/2 = 0 \bigoplus 1$, and occurs because, in the same way that particles can be classified by their transformation properties under rotations (or total spin), excitations in a topological theory are associated with representations of a (quantum) group\footnote{In many cases the representation theory of the group is deformed in the topological theory, such that there are only a finite number of representations even for continuous groups}.  When the group in question is non-abelian, most excitations will not be simple currents.  The general technique employed here to construct the Hamiltonian (\ref{Eq_HFull}) by adding a term that pair-creates vortices on adjacent plaquettes still applies in such cases.  Condensing bosons with non-Abelian fusion rules  will lead to a different critical behavior, which is not equivalent to that of any statistical mechanical model that the authors are aware of.   The study of these transitions is undoubtedly a rich  subject for further study; one interesting example is discussed in Ref. \onlinecite{Gilsetal}.

The second restriction we have imposed here is to consider only achiral condensates, by condensing plaquette violations in the lattice model.  An obvious question is whether chiral condensates (or condensates of vertex violations, in the lattice model) can also occur.  From the purely topological viewpoint there is no obstruction to forming these\cite{TSB}, provided that the excitation to be condensed is a boson.  (This is always the case for achiral excitations, but need not be for their chiral cousins).  Further, the critical theory will again be of the transverse-field Potts type if the condensed boson is a simple current.  In the lattice model, however, only when there is no splitting is it clear that operators for all excitations in the condensed phase can be constructed.  In the absence of splitting, the Chain-Mail\cite{CH} formulation of the partition function can be used to show that the final theory is dual to an achiral condensate, and the resulting duality mapping between the string operators gives an explicit representation of all excitations in the final theory.  When splitting occurs, this duality fails and there appear to be no conserved quantum numbers to differentiate the split particle types, suggesting that the final topological phase may not be fully realizeable by the lattice model.  We will discuss these results in more detail in a future work.

The task of fully categorizing the possible phase transitions and critical theories between phases of different topological order remains a source of many open questions.  The solvable Levin-Wen\cite{LW} models considered here provide a useful framework in which to rigorously study such questions; since both topological order and properties of the critical theory are relatively universal, conclusions drawn from the lattice model also apply to real physical systems exhibiting the desired topological characteristics, where these exist.

\appendix

\section{ Magnetic quasi-particle operators} \label{QPApp}

Here we discuss in more detail the form of operators creating magnetic quasi-particles which do not correspond to simple currents.
The precise form of the general magnetic quasi-particle operators is complicated by the fact that, in the uncondensed phase, if the label $i$ does not correspond to a simple current, there is no {\it purely} magnetic excitation associated with $i$.  Rather, what we will call the magnetic $i$ excitation is in fact a specific superposition of excitations which all carry magnetic flux $i$, but also carry electric flux $j \in i \times i$.

The reason for this is that the fundamental excitations in our Levin-Wen model are not electric and magnetic fluxes as would be the case for a lattice gauge theory, but the sources of the right- and left- handed Chern-Simons fields.  As explained in Ref. \onlinecite{CH}, the right-handed sources are precisely the electric sources described above.  The left-
handed sources $i_L$, however, carry both the electric charge $\overline{i}$ and the magnetic flux $i$.  The best approximation to a purely magnetic excitation in this case is the achiral source $i_L \times i_R$, which has magnetic flux $i$ and electric flux $i \times \overline{i} = Id + ... $.  In general the individual electric flux labels on the right-hand side are not conserved along the length of a particular string, so that the entire superposition is required to construct the appropriate quasi-particle operator.

\begin{center}
\begin{figure}[h]
\includegraphics{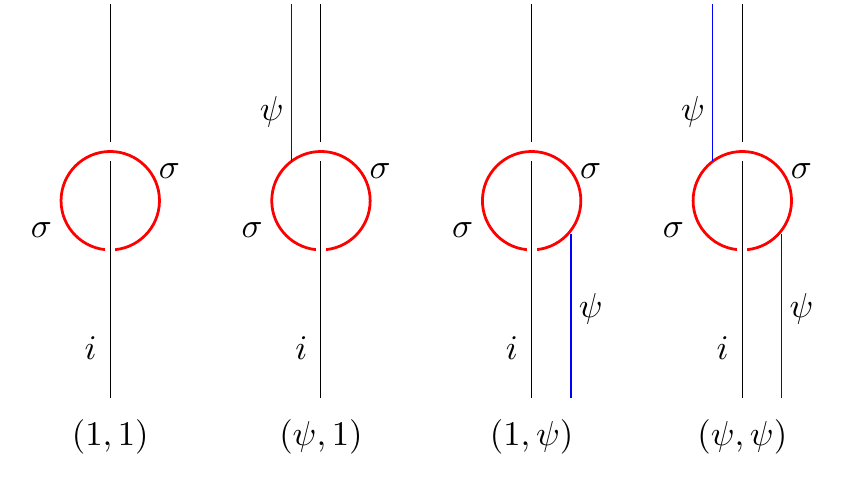}
\caption{ \label{QP12Fig} The  $\sigma$ vortex in the doubled Ising anyon theory consists of a pair of electric sources (one right- and one left- handed).  Its action on an edge is given by a phase (depicted here by the red ring) every time the string crosses between plaquettes, together with an electric component which raises the edge label by $Id$ or $\psi$.  This gives the four possibilities shown here each time the string operator crosses between plaquettes.  The labels $(Id,Id), (\psi, Id), (Id, \psi)$, and $(\psi, \psi)$ denote the associated electric flux on the upper and lower sides of the crossing.  }
\end{figure}
\end{center}

To illustrate how this works in practice, we consider the  $\sigma$ vortex in the doubled Ising anyon theory.  (This is essentially the same as the spin-$1/2$ vortex in a doubled $SU(2)_2$ Chern-Simons theory, though the signs in this case are somewhat simpler to track).  The operator is represented diagramatically in Fig. \ref{QP12Fig}; it acts on the edge labeled $i$ in the Figure according to
\ba
\hat{s}_{(1/2, 1/2)} |i \rangle & =& \frac{1}{2} \left(  S_{1/2, i} | i \rangle +  (-1)^{n_\psi} S_{1/2, i} | i \times \psi \rangle \right) \n
&&+ \frac{1}{\sqrt{2}} \left ( e^{- \frac{ i \pi}{4} } F_L(i) + e^{\frac{ i \pi}{4} }F_R(i)  \right ) \delta_{i \sigma} |\sigma \rangle \n
\ea
where $F_{L,R} = \pm 1$ are coefficients which depend on the labels of adjacent edges at the left (L) and right (R) ends of the link respectively.  This action can be decomposed into the action of the four channels shown in the figure, as shown in Table \ref{StringTab}.  The coefficients $F_{L,R}$ come from the action of the $\psi$ tail on the left- and right- vertices.

\begin{center}
\begin{table}
\begin{tabular}{|c|c|c|c|c|}
\hline
$i$ & $(Id,Id)$ & $(Id, \psi)$ & $(\psi, Id )$ & $(\psi, \psi)$ \\
\hline
$1$ & $\frac{1}{\sqrt{2}} 1$ & $0$ & $0$ & $\frac{1}{\sqrt{2}} \psi$ \\
$\psi$  & $-\frac{1}{\sqrt{2}} \psi $& $0$ & $0$ & $\frac{1}{\sqrt{2}} 1 $ \\
$\sigma$ &  $0$& $\frac{1}{\sqrt{2}}e^{ \frac{- i \pi}{4} }  \sigma$  & $\frac{1}{\sqrt{2}}e^{ \frac{ i \pi}{4} }  \sigma$&  $0$  \\
\hline
\end{tabular}
\caption{ \label{StringTab} Action of the four possible combinations of fusion channels for the $\sigma_L \times \sigma_R$ excitation (the $\sigma$ vortex).  Here we have omitted any factors associated with labels on adjacent edges.  }
\end{table}
\end{center}

It is important to note that if we keep only the $(Id,Id)$ and/or $(\psi, \psi)$ fusion channels, the action of this operator on a given edge is unchanged if we simultaneously act with the operator $(-1)^{n_\sigma}$ which creates a pair of $\psi$ vortices.  This means that keeping only the $(Id,Id)$ and $(\psi,\psi)$ channels adds an indefinite number of $\psi$ vortices to each plaquette.  In the condensed phase, where $\psi$ vortices are in any case not conserved, this does  not affect  the energy of states this operator creates; in the uncondensed phase, however, if the $\sigma$ vortices at the end-points of the string are more than one plaquette apart then the state is not a single eigenstate, but rather a superposition over all eigenstates with some number of $\psi$ vortices on the intervening plaquettes.  As a consequence this operator is not topological, since the location of these possible $\psi$ vortices on the lattice depends on its trajectory.   Once we include the effect of the $(Id, \psi)$ and $(\psi, Id)$ channels, which do not annihilate $\sigma$ labels, this is no longer the case and the operator creates only a pair of $\sigma$ vortices at each of its endpoints.

%One might ask what happens if, at a given link crossing in the lattice, we switch from an operator which includes all four possibilities to one which includes only the $Id$ channel.  This corresponds to changing from an operator with an unobservable string to an observable one, in the solvable limit.   Since the operator can switch between the $Id$ and $\psi$ channels only on edges labeled $\sigma$, the relevant question is what has happened to the missing $\psi$ flux that would normally have some probability of emenating from the last $\sigma$ edge that the operator crossed.  The answer is that the operator has some probability of creating a $\psi$ vertex violation at one of the two vertices bordering this edge.  We interpret this as meaning that the pure $\sigma$ vortex creation operator ends on this edge, and that the extra $\sigma$ operators in the $Id$-fusion channel are from another, non-topological, string operator which also has some amplitude to create $\psi$ vortices, as described above.

The case for general magnetic sources is similar: we find that only when all of the fusion channels of $i \times \overline{i}$ are included as electric source lines is the string operator topological.  If some of these are omitted, then the operator has some probability of creating extra vortices on the plaquettes separating the two $i$ vortices, and hence is not topological as the number of possible violations scales with the separation between the string's endpoints.

\section{Splitting in general condensates} \label{SplittingApp}

Having detailed the form of composite operators $a_L \times a_R$ in the previous section, we now present a proof that  condensing  an achiral simple current will always lead to the correct splitting of string operators in the condensed phase.
It is useful to begin with the example discussed above, and consider the splitting of the $\sigma_L \times \sigma_R$ particle in the doubled Ising theory.

From the action of the possible fusion channels of the $\sigma$ vortex in Table \ref{StringTab}, it is easy to see that when the edge label $\sigma$ is eliminated in the topological limit of the condensed phase, the $(Id,Id)$ and $(\psi, \psi)$ channels do not mix.  One way to understand why this happens is to note that before condensation these string operators were non-topological because they created a trail of plaquettes containing a superposition of no vortex and the $\psi$ vortex; hence their average energy depends linearly on the string length in this r{\'e}gime.  After condensation, however, the presence of a $\psi$ vortex does not change the energy of a state, and these become legitimate topological string operators.

\begin{center}
\begin{figure}[h]
\includegraphics[height=1in]{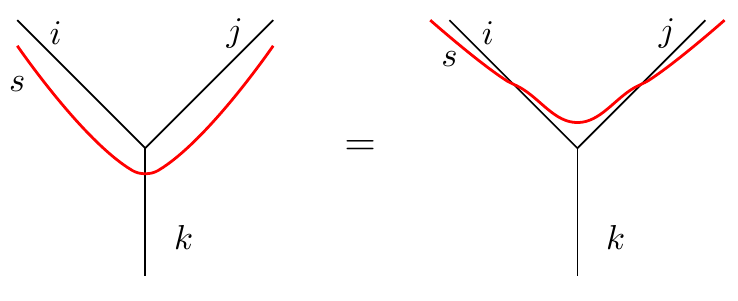}
\caption{\label{SlideFig} String operators creating topological excitations can slide freely over vertices, without altering the state of the system.  String operators with this property are said to obey the {\it hexagon relation}\cite{LW}; the corresponding excitations are topological because their energy is independent of the trajectory which the string takes through the lattice.  String operators that are not topological leave behind a trail of excitations.  After condensation, some achiral particles may split -- meaning that sub-sets of the possible edge labels associated with a particle $a_L \times a_R$ separately obey the hexagon relation.  This happens because the trail of excitations that these string operators leave behind are vortices which have condensed.}
\end{figure}
\end{center}

This basic argument can be generalized to other condensates of achiral simple currents.  To do so, we exploit the fact that topological excitations of the (solvable) Levin-Wen Hamiltonian describing the condensed phase must obey the {\it hexagon relations}.  From the point of view of the lattice model, these simply state that any string operator creating a topological excitation can slide freely over vertices, as shown in Fig. \ref{SlideFig}.  As in the case of the $\sigma_L \sigma_R$ particle, string operators which do not satisfy the hexagon relation are not topological because they leave a trail of excitations along their length, rather than just at their endpoints.  In some cases, however, this trail of excitations consists entirely of the vortices that we condense.  When this happens the corresponding string operators {\it are} topological in the condensed phase.  Our purpose here will be to show that this process accounts exactly for the splitting expected from the TSB criterion.  Readers should note that to do this we will make reference to tensors $F$ (the $6j$- symbols) and $R$ (the universal $R$ matrix) which are determined by the choice of topological order.  We will not explain their meaning here, but a useful introduction can be found in Refs. \onlinecite{Preskill,BondersonAnnals,BondersonThesis,CH,KitaevVeryLongPaper}.

\begin{widetext}
%Recall that such an operator will split if and only if $a\times \phi^k = a$ for some $1<k < Q$.
Before condensation, in the doubled Chern-Simons theory all string operators have the form $\hat{a}_L \times \hat{b}_R$.   These can be expressed in the form of a phase operator $\hat{R}_{ab}$ acting on each edge that the string crosses (diagramatically a ring  labeled $(a,b)$  encircling the crossed edges, as shown in Fig. \ref{QP12Fig}), and operators $\hat{s}_{X} \hat{s}_{Y}$ which raise or lower the electric flux on each edge by $b_i$ and $b_f$ to either side of the ring.  We will focus on the configuration shown in Fig. \ref{SlideFig}, in which the string operator crosses over a pair  of edges.  For example, $\hat{a}_L \times \hat{a}_R$ has the form:
\be
\hat{a}_R \otimes \hat{a}_L = \hat{R}_{aa} \sum
%_{b \in \mbox{ Possible fusion products of }a \otimes a }
F^{a^* a 0}_{a^* a b} \hat{b} \ \ \ .
\ee
where $F^{a^* a 0}_{a^* a b}$ is a coefficient ($6j$ symbol) dependent on the labels $a$ and $b$.
The reason that this particle satisfies the hexagon relation is that
since the sum runs over {\em all}  values of $b$ allowed by fusion, we may `undo' the process of making the composite operator, to depict this operator as two separate strings, labeled $\hat{a}_R$ and $\hat{a}_L$ respectively.  In the doubled Chern-Simons models, it is easy to show that the string operators $\hat{a}_{L},\hat{a}_{R}$ do obey the hexagon relation\cite{CH}.  Graphically, we may depict the situation as follows:
\begin{center}
\includegraphics[width=5.5in]{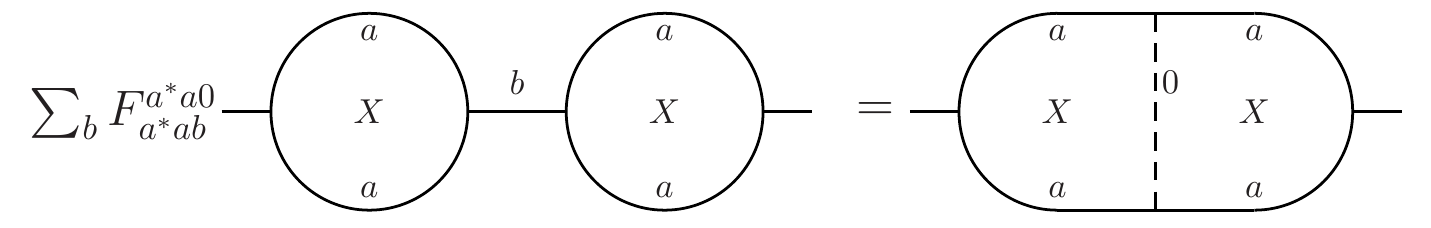}
\end{center}
Here $X$ indicates that each $a$ loop is encircling a labeled edge in the lattice, as is appropriate for the phase operator $R_{aa}$.  After re-expressing the operator locally in terms of the two strings $\hat{a}_L$ and $\hat{a}_R$, the resulting operator may be pulled over a vertex at which the two edges $X$ join.  This ensures that the hexagon relation (Fig. \ref{SlideFig}) is obeyed.

Before condensation, only when the dashed line carries the label $0$ can we pull the $a$-loop  on the right-hand side over a vertex, and hence there is only one topological string operator associated with $\hat{a}_L \times \hat{a}_R$.  After condensation, however, any combination of coefficients on the left-hand side which results in the dashed line carrying powers of $\phi$ gives an operator that satisfies the hexagon relation, and hence a valid topological quasi-particle.  Here we assume that there are no other labels in the category which braid trivially with all of the deconfined edge labels; hence we wish to find all linearly independent sets of coefficients $C(b)$ such that:
 \begin{center}
\includegraphics[width=6in]{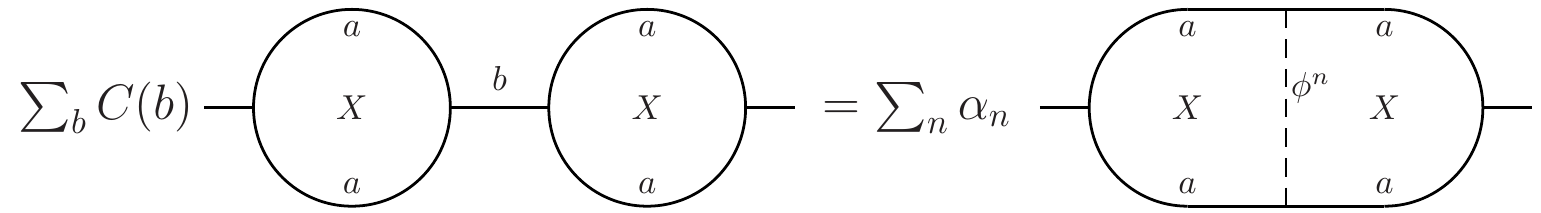}
\end{center}
where the coefficients $\alpha_n$ are arbitrary.
\end{widetext}

Our task now is to count the number of possible linearly independent sets of coefficients $C(b)$ will have the property that
\be \label{Eq_QPAllow}
\sum_b C(b) F^{a a^* b^*}_{a a^* d} = \sum_{n=0}^{Q-1} \alpha_n \delta_{d, \phi^n} \ \ \ .
\ee
 This will give us the number of different particle types, as each dimension in the vector space of possible solutions to (\ref{Eq_QPAllow}).  Clearly, if $\phi_n \times a \neq \overline{a}$, then $\alpha_n =0$ as the diagram on the right is not consistent with the fusion rules of the theory.  This leaves $Q/k$ possible linear combinations on the right-hand side of Eq.~(\ref{Eq_QPAllow}), corresponding to the $Q/k$ independent particle types.  

To show this explicitly, we will identify the $Q/k$ linearly independent sets of coefficients $C(b)$.  
We begin with the choice $C(b) = \frac{\sqrt{ \Delta_b}}{ \Delta_a} = F^{a^* a 0}_{a^* a b}$.  With this choice, we have:

\begin{center}
\includegraphics[width=3.5in]{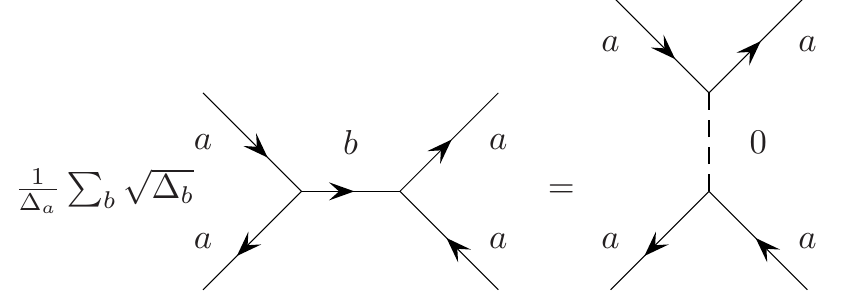}
  \end{center}
and hence $C(b) = \frac{\sqrt{ \Delta_b}}{ \Delta_a} $ is certainly one of the allowed sets of coefficients.
Next, we consider diagrams of the form:

\begin{center}
\includegraphics[width=3.5in]{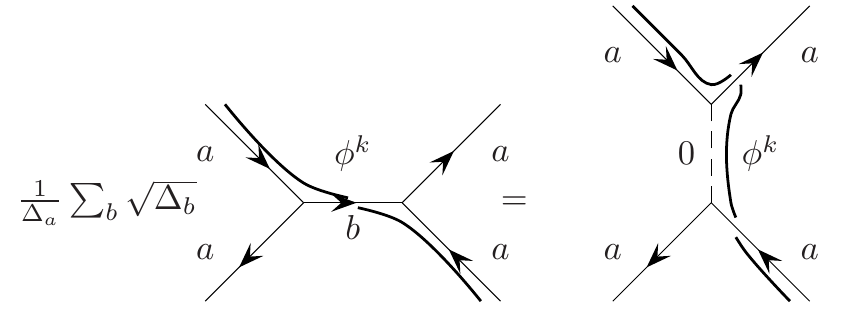}
  \end{center}
If $\phi^k \times a = a$, then the four external edges of the diagram are still labeled $a$.  Further, the vertical line on the right-hand side clearly carries the label $\phi^k$, and hence the particle depicted obeys the hexagon relation in the condensed phase.
Specifically, the identity above is equivalent to:
\begin{center}
\includegraphics[width=3.5in]{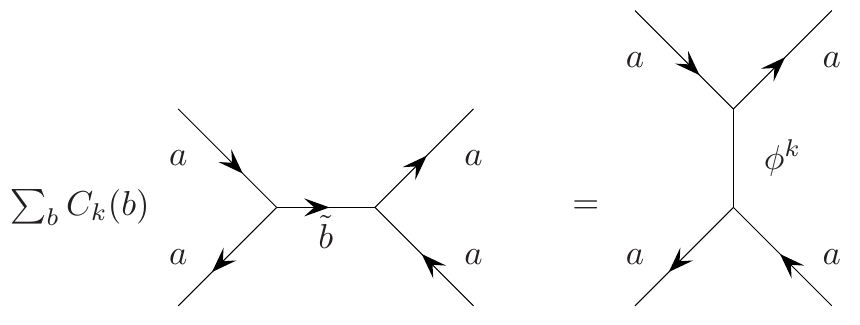}
\end{center}
with $\tilde{b} = \phi^k \times b$, and
\be
C_k(b) = \frac{1}{\Delta_a}  \sqrt{\Delta_b}   F^{(\phi^k)^* a  a^*}_{a^* b^* \tilde{b}}
	 F^{\phi^k a a^*}_{a^* b \tilde{b}^*} R^{ \phi^k b}_{\tilde{b}} \ \ \ .
\ee
We thus have a candidate choice of $C_{n k}$ for each of the $Q/k$ possible values of $n$.  These must all be linearly independent, since the vertical lines of the diagrams on the right carry different powers of $\phi$.  Any other choice of coefficients either gives a superposition of these $Q/k$ possibilities, vanishes, or produces a quasi-particle type that does not obey the hexagon relation.  In particular, $a \times a$ contains $\phi^j$ only if $\phi^j \times a =a$, so that no other powers of $\phi$ may appear on the right-hand side.

  It is useful to express this statement in matrix form.  We can write
\begin{center}
  \begin{tabular}{cccc}
 $ C_{0}(b_1) $ & $ C_0 (b_2) $ & $ ... $& $C_0 (b_r)$ \\
   $ C_{1}(b_1) $ & $C_1 (b_2) $ & $ ...$ & $C_1 (b_r) $\\
    \vdots & \vdots &  \vdots &  \vdots \\
   $  C_{Q/k}(b_1) $ & $ C_{Q/k} (b_2)$ &$ ... $& $C_{Q/k} (b_r)$ \\
\end{tabular}
\end{center}
where $r$ is the number of possible fusion outcomes of $a \otimes a$, and in particular $r \geq Q/k$ as $a \otimes a = Id + \phi^k + ... + \phi^{N-k} + ... $.  The $Q/k$ orthogonal linear combinations of coefficient vectors which can be formed from these ensure the existence of $Q/k$ distinct string operators.  In general these string operators will generate linear combinations of the true quasi-particle types in the theory; to identify these requires additional physical input (such as their self-braiding statistics).  

It is useful to illustrate how the above counting procedure works for the case of \su.  Here the coefficients are
\ba
C_0(j) &=& \frac{1}{\sin \pi/(k+2) } \sin \left ( \frac{ (2j+1 )\pi}{k+2 } \right ) \n
C_1(j)& =&(-1)^{j } \frac{1}{\sin \pi/(k+2) } \sin \left ( \frac{ (2j+1 )\pi}{k+2 } \right )
\ea
Thus we see that there are indeed two linearly independent combinations of coefficients which will produce the desired result.  The particle types are given by taking the sum and difference of these, to obtain either all even integer or all odd integer spins on the edges.  (For example, in $SU(2)_2$, this gives the quasi-particle operators $(\frac{1}{2}, \frac{1}{2})_1$ and $(\frac{1}{2}, \frac{1}{2})_0$).  As we have seen by direct computation, these are indeed the two expected particle types.

{\bf Acknowledgements:} 
SHS acknowledges funding from an SFI ETS Walton fellowship.  JKS was supported by Science Foundation Ireland Principal
Investigator award 08/IN.1/I1961.  The authors are grateful to S.L. Sondhi, R. Stinchcombe, M.A. Levin, D. Huse, and B. Halperin for helpful discussions, and for the hospitality of the Aspen Center for Physics.

\bibliography{TSBBib}

\end{document}